\documentclass[final,3p,times]{elsarticle}

\usepackage{amssymb}
\usepackage[utf8]{inputenc}
\usepackage{subfig}
\usepackage{siunitx}
\usepackage{amsmath,color}
\graphicspath{{./bilder/}}
\newcommand{\coma}{\>\text{,}}
\newcommand{\point}{\>\text{.}}

\newcommand{\legline}[1]{\raisebox{-0.1cm}{\protect\includegraphics{./line#1.pdf}}}

\newcommand{\legpoint}[1]{\raisebox{-0.1cm}{\protect\includegraphics{./point#1.pdf}}}

\usepackage{multirow}
\usepackage{booktabs}
\usepackage{blindtext}

\journal{CMAME\qquad \texttt{{\normalfont \textcolor{red}{http://dx.doi.org/10.1016/j.cma.2018.11.020 (added on Nov. 30, 2018)}}}}

\begin{document}

\begin{frontmatter}



\title{Phase-Field Modelling of Interface Failure in Brittle Materials}


\author[label1]{Arne Claus Hansen-Dörr}
\author[label3]{René de Borst}
\author[label1]{Paul Hennig}
\author[label1,label2]{Markus Kästner\corref{cor1}}

\address[label1]{Institute of Solid Mechanics, TU Dresden, Dresden, Germany}
\address[label2]{Dresden Center for Computational Materials Science (DCMS), TU Dresden, Dresden, Germany}
\address[label3]{University of Sheffield, Department of Civil and Structural Engineering, Mappin Street, Sir Frederick Mappin Building, Sheffield S1 3JD, UK}
\cortext[cor1]{markus.kaestner@tu-dresden.de}

\begin{abstract}
A phase-field approach is proposed for interface failure between two possibly dissimilar materials. The discrete adhesive interface 
is regularised over a finite width. Due to the use of a regularised crack model for the bulk material, an interaction between the 
length scales of the crack and the interface can occur. An analytic one-dimensional analysis has been carried out to quantify this 
effect and a correction is proposed, which compensates influences due to the regularisation in the bulk material. For 
multi-dimensional analyses this approach cannot be used straightforwardly, as is shown, and a study has been undertaken to 
numerically quantify the compensation factor due to the interaction. The aim is to obtain reliable and universally applicable 
results for crack propagation along interfaces between dissimilar materials, such that they are independent from the regularisation 
width of the interface. The method has been tested and validated on three benchmark problems. The compensation is particularly 
relevant for phase-field analyses in heterogeneous materials, where cohesive failure in the constituent materials as well as 
adhesive failure at interfaces play a role.
\end{abstract}


\end{frontmatter}


\section{Introduction}
\label{sec:intro}

The functionality of engineering structures may be compromised by cracks, and there has been a long-standing interest to predict 
crack initiation and propagation, i.e. the location at which a crack nucleates, under which conditions it propagates, and its 
propagating direction.

The work of {Griffith} \cite{griffith_phenomena_1921} is a landmark contribution in the understanding of fracture. He introduced 
an energetic criterion to assess whether crack growth would occur or not. Herein, a central concept is the fracture toughness, 
which is an energetic threshold. Often, crack growth is a highly transient process, incorporating complex mechanisms like crack nucleation, propagation, branching and possibly crack arrest, i.e. such analyses can only be carried out 
numerically. Computational approaches can roughly be divided into two categories, namely methods in which the crack is represented 
in a discrete manner, and those in which a diffuse, or continuum,  representation is employed, 
e.g.~\cite{de_borst_discrete_2004}.

Herein, we will focus on a particular method within the class of diffuse representations, namely the phase-field approach to 
fracture. %, which recently has become en vogue in computational settings. 
The phase-field model covers crack nucleation and crack 
propagation, yields qualitatively good results for homogeneous materials and is appealing due to its conceptual simplicity. Indeed,
phase-field models are a powerful way of modelling cracks, especially when it comes to three-dimensional problems. Moreover, they 
can straightforwardly handle issues that may be complicated in discrete crack analyses, such as a priori unknown crack paths, 
crack arrest and branching. In this contribution we will show, however, that modelling adhesive failure along sharp interfaces 
is more complex and requires additional numerical treatment.

Starting from a Griffith approach, Francfort and co-workers~\cite{francfort_revisiting_1998} introduced an integral crack surface 
energy formulation where the total energy takes a minimum for the correct crack path. Subsequently, this formulation was 
regularised, resulting in a diffuse crack representation~\cite{bourdin_numerical_2000,bourdin_variational_2008}. 
The crack is now no longer described by a physical crack opening, but is rather represented by a scalar field over the entire 
domain. This, in principle, avoids the need for remeshing and is appealing for complex fracture processes. The auxiliary scalar
variable, often referred to as the phase-field variable, regularises the boundary value problem and distributes the discrete crack 
over a finite width. In an extension of the above approach for brittle fracture to quasi-brittle and ductile fracture, Verhoosel
and de Borst~\cite{verhoosel_phase-field_2013} incorporated the cohesive-zone model in the phase-field approach.% by describing the location of the discontinuity through a phase-field, while introducing an additional independent variable, the displacement jump over the discontinuity as input parameter for evaluating the traction-relative displacement relation.

Engineering materials are often composed of several components, e.g. reinforced concrete, or fibre-reinforced composites, which 
consist of fibres and a matrix, or laminates with different plies. Often, the interfaces in such heterogeneous material systems
are the weak spots where fracture initiates. A complete failure analysis of such composite materials or structures therefore
requires that cracks can propagate within the matrix with an a priori unknown crack path, but also along interfaces between
two dissimilar materials. Typically, the fracture toughness of an adhesive interface, say $\mathcal{G}_\text{c}^\text{int}$,
is different from that in the bulk material, $\mathcal{G}_\text{c}^\text{bulk}$. 

Schneider et al.~\cite{schneider_phase-field_2016} have presented a multiphase-field model which is capable of describing cracks 
within grains as well as along grain boundaries. The modification of the surface energy between the bulk materials allows for 
different interface properties.  In a hybrid approach, Paggi~et~al.~\cite{paggi_revisiting_2017} incorporated a sharp interface 
by combining a phase-field model for brittle fracture with a cohesive-zone model for the interface. For a brittle, inclined 
interface, they achieved results comparable to linear elastic fracture mechanics derived by He and 
Hutchinson~\cite{he_crack_1989}. Nguyen~et~al.~\cite{nguyen_phase-field_2016} extended the standard cohesive zone approach by introducing an interface regularisation similar to the crack phase-field, where the displacement jump over the interface takes a regularised form. 

This work extends the findings of {Hansen-Dörr}~et~al.~\cite{hansen-dorr_numerical_2017}, who presented a method to incorporate an 
adhesive interface in the bulk material using a phase-field model for brittle fracture~\cite{miehe_phase_2010}. This qualitative 
study showed a significant interaction between the surrounding bulk material and the diffuse interface due to the different length 
scales of the crack and the interface. Indeed, the critical energy release rate at which the crack propagates is not only governed 
by the fracture toughness of the interface, but also by the fracture toughness of the bulk material. 

Our aim is to correct the influence of the bulk material on crack propagation along an interface by modifying the fracture 
toughness of the interface. The testing environment to investigate the proposed modification should be such that:
\begin{itemize}
	\item There is a steady, controllable crack growth along the interface; 
	\item The method to calculate the energy release rate can be used to compare the results to the predefined values;
	\item There is an efficient spatial discretisation of the regularised crack surface.
\end{itemize}
The first requirement is met using a so-called surfing boundary condition, cf.~\cite{hossain_effective_2014,kuhn_discussion_2016}. 
The second issue is addressed using the concept of configurational forces which exploits a generalised force acting on the crack 
tip~\cite{kuhn_discussion_2016,kuhn_energetic_2011}. An accurate and efficient representation of the steep gradient of the 
regularised crack is obtained using isogeometric analysis, cf. \cite{borden_phase-field_2012}, with local refinement~\cite{hennig_bezier_2016,hennig_adaptive_2016}. 

The paper is structured as follows. Section~\ref{sec:theory} gives a concise review of the phase-field model for brittle fracture.
Section~\ref{sec:interface1D} introduces the idea of a diffuse interface and discusses the interaction of both length scales. An 
exact solution is derived for a proper compensation in a one-dimensional case. The next section shows that the one-dimensional 
approach cannot be taken over to two-dimensional configurations in a straightforward manner and presents a numerical study that
provides compensation factors for two-dimensional cases.  Section~\ref{sec:numeric} presents numerical examples and a comparison 
with analytical results. Concluding remarks are drawn in Section~\ref{sec:conclusion}.
%%%%%%%%%%%%%%%%%%%%%%%%%%%%%%%%%%%%%%%%%%%%%%%%%%%%%%%%%%%%%%%%%%%%%%
\begin{figure}[t]
	\centering
	\subfloat{\includegraphics{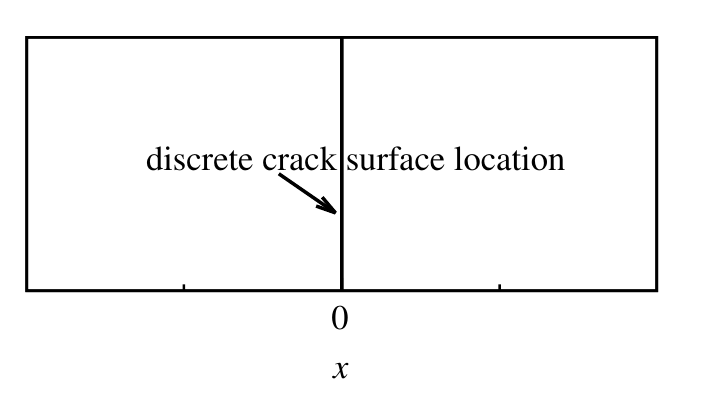}}
	\subfloat{\includegraphics{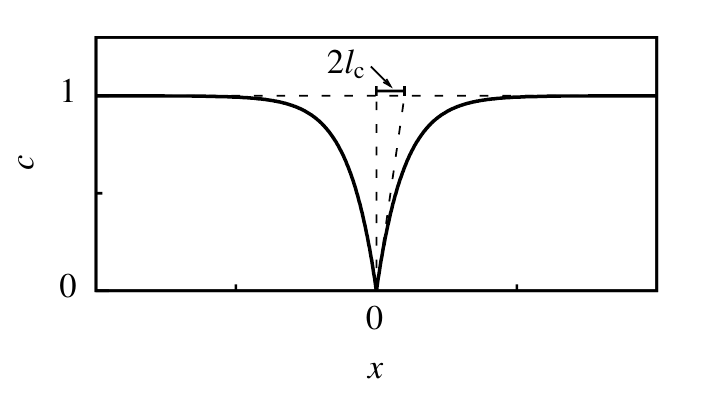}}
	\caption{The discrete crack surface depicted on the left is regularised using an exponential function. The resulting distributed discontinuity, regularised using the length scale parameter $l_\text{c}$, is shown on the right.}
	\label{fig:smearedcrack}
\end{figure}

\section{Phase-field Model for Brittle Fracture}
\label{sec:theory}
The phase-field approach introduces an additional scalar field $c$, which in a continuous manner separates fully intact material
($c=1$) from fully broken material ($c=0$), and is coupled to the mechanical part of the boundary value problem. 
Figure~\ref{fig:smearedcrack} illustrates the regularisation. The originally sharp crack $\Gamma$ is regularised through a 
phase-field $c$, yielding a steep gradient near the original crack surface. Following~\cite{bourdin_variational_2008}, 
the total energy of a brittle, linear elastic (cracked) domain reads
\begin{equation}
\label{eq:energy}
\Psi=\Psi^\text{el}+\Psi^\text{c}=\int_{\Omega}\psi^\text{el}+\psi^\text{c}\,\text{d}v=\int_{\Omega}(g(c)+\eta)\cdot\psi^\text{el}_++\psi^\text{el}_-+\underbrace{\frac{\mathcal{G}_\text{c}}{4l_\text{c}}\left\lbrace\left(1-c\right)^2+4l_\text{c}^2 c_{,i}c_{,i}\right\rbrace}_{\text{crack/dissipated energy}} \,\text{d}v\coma
\end{equation}
where $l_\text{c}$ is the characteristic length scale that governs the width of the crack phase and 
$(\bullet)_{,i}=\partial(\bullet)/\partial x_i$. The summation convention applies. For the specific elastic \textsc{Helmholtz} 
free energy $\psi^\text{el}=\tfrac{1}{2}\boldsymbol{\varepsilon}:\boldsymbol{E}:\boldsymbol{\varepsilon}$ a split is carried out
into a tensile~$\psi^\text{el}_+$ and a compressive~$\psi^\text{el}_-$ part to avoid cracking under compressive 
stresses~\cite{miehe_thermodynamically_2010}. The degradation function $g(c)=c^2$ fulfils the condition that the driving force of 
the phase-field $c$ vanishes when $c=0$. The fracture toughness $\mathcal{G}_\text{c}$ is the material parameter within the energetic crack growth
criterion of Griffith~\cite{griffith_phenomena_1921}. The residual stiffness $0<\eta\ll1$ prevents numerical instabilities.

The Euler-Lagrange equations, which describe the coupled problem, can be derived in a variational 
manner~\cite{miehe_phase_2010,miehe_thermodynamically_2010}. Neglecting volume forces, they read
\begin{align}
\label{eq:momentum}
\sigma_{ij,i}&=0\quad\text{and}\\
\label{eq:pf}
\left[1-c\right]+4l_\text{c}^2c_{,ii}&=\frac{2g^\prime(c)l_\text{c}}{\mathcal{G}_\text{c}}\psi_+^\text{el}\coma\\
\text{with}\quad\sigma_{ij}&=g(c)\frac{\partial\psi_+^\text{el}}{\partial\varepsilon_{ij}}+\frac{\partial\psi_-^\text{el}}{\partial\varepsilon_{ij}}
\end{align}
subject to the boundary conditions
\begin{align}
\sigma_{ij}n_i=\bar{t}_j& \quad\text{on $\partial\Omega_t$}\coma\\
u_i=\bar{u}_i& \quad\text{on $\partial\Omega_u$}\coma
\end{align}
of the momentum equation~\eqref{eq:momentum}, and subject to 
\begin{equation}
c_{,i}n_i=0\quad\text{on $\partial\Omega$}\coma
\end{equation}
of the phase-field equation~\eqref{eq:pf}, where $\partial\Omega=\partial\Omega_t\cup\partial\Omega_u$ is the boundary of the body 
and $\emptyset=\partial\Omega_t\cap\partial\Omega_u$.

Irreversibility of the crack evolution can be enforced in different ways. In the damage mechanics interpretation of Miehe and co-workers~\cite{miehe_phase_2010,miehe_thermodynamically_2010} the phase-field
variable is similar to a gradient damage model with a smooth transition between the intact and fully broken state. While this 
damage-like interpretation of the phase-field approach to brittle fracture is appealing -- for a further discussion on similarities 
and differences the reader is referred to Reference~\cite{de_borst_gradient_2016} -- it turns out that the interpretation of the 
phase-field variable as a history parameter may compromise the convergence of the functional that describes the diffuse crack to 
that which describes the discrete crack~\cite{vignollet_phase-field_2014,may_numerical_2015,linse_convergence_2017}, and that fixing the phase-field variable when it is very close to the value that indicates complete local failure, 
e.g.~\cite{kuhn_continuum_2010}, is to be preferred. For this reason the latter approach, also known as fracture-like constraint 
\begin{equation}
c = \begin{cases}
c & c> c_\text{th} \\
0 & c< c_\text{th}
\end{cases}\coma
\end{equation} 
is used here, with the threshold $c_\text{th}=0.01$. In other words, as soon as the phase-field reaches a value below the very small threshold $c_\text{th}$, a Dirichlet boundary condition $c=0$ is applied at the corresponding node in the finite element framework.

The weak form is discretised using locally refined Truncated Hierarchical B-splines (THB-splines)~\cite{hennig_bezier_2016}. 
This allows for efficient computations with a high resolution of the steep gradient in regions where the crack develops. In the framework of isogeometric analyses (IGA), the control points of the splines, which are used to interpolate the field quantities, adopt the role of the nodes which are present in classical finite element analyses. The resulting non-linear equations are solved using a staggered scheme~\cite{miehe_phase_2010}. Iterations are carried out within this 
staggered scheme to ensure convergence of both fields at each load level. Displacement control has been used in all examples.

%%%%%%%%%%%%%%%%%%%%%%%%%%%%%%%%%%%%%%%%%%%%%%%%%%%%%%%%%%%%%%%%%%%%%%
\section{Phase-field Modelling of Interface Failure: The one-dimensional problem}
\label{sec:interface1D}

\begin{figure}[t]
	\centering
	\begin{minipage}{0.55\textwidth}
		\centering
		\subfloat[DCB geometry with diffuse interface]{
			\label{fig:prelimdcb_geom}
			\includegraphics[trim={0cm 0cm 0cm 0cm},clip]{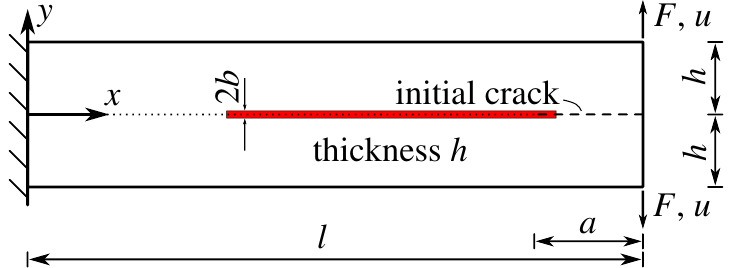}
		} \\ \vspace{0.5cm} \footnotesize
		\begin{tabular}{l| l}
			\legline{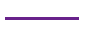} $b=\SI{62.5}{\micro\meter} $& \legline{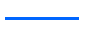} $b=\SI{15.625}{\micro\meter} $ \\
			\legline{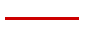} $b=\SI{31.25}{\micro\meter} $ & \legline{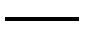}  $b=\SI{7.8125}{\micro\meter} $\\
		\end{tabular}
	\end{minipage}
	\begin{minipage}{0.4\textwidth}
		\centering
		\subfloat[Force-displacement curves ]{
			\label{fig:prelimdcb_fd}
			\includegraphics[trim={0cm 0cm 0cm 0cm},clip]{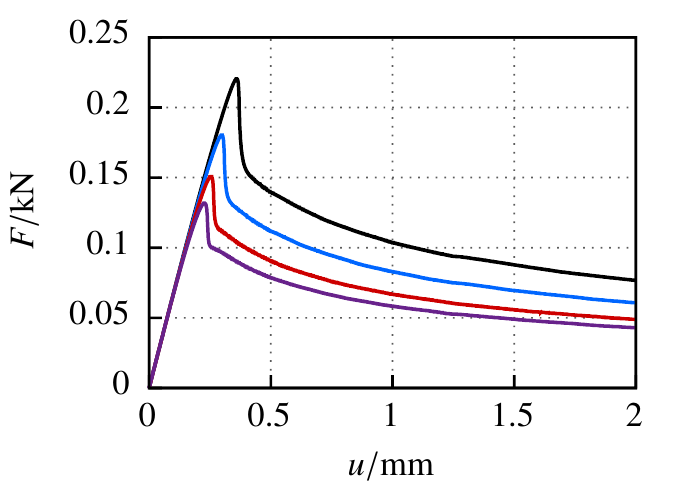}
		}
	\end{minipage}
	\caption[Interface failure]{Interface failure: (a) Domain ($l=10a=20h=\SI{10}{\milli\meter}$) and boundary conditions for the preliminary DCB test and red region with reduced $\mathcal{G}_\text{c}^\text{int}$. (b) Influence of the choice of the interface half-width $b$ on the material strength, while keeping $l_\text{c}=\text{const}=\SI{50}{\micro\meter}$.}
	\label{fig:prelimdcb}
\end{figure} 

\begin{figure}[t]
	\centering
\includegraphics{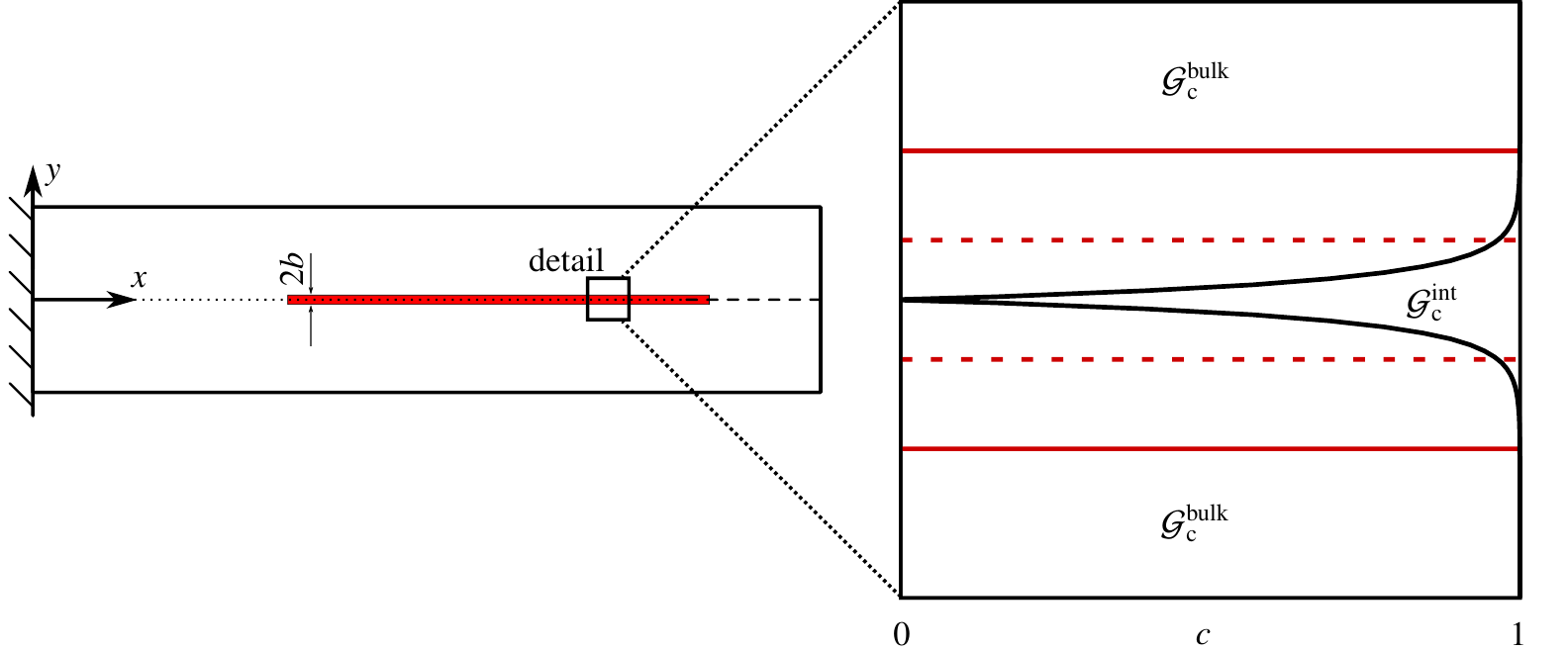}
	
	\caption[Interface failure]{On the right, the detail describes a fully developed crack within the interface in the DCB. The solid red lines mark the original interface of width $2b$ as shown on the left. The fracture toughnesses of the interface and the bulk material differ. Reducing the interface width (dashed red lines) has an influence on the dissipated energy since the fracture toughness is heterogeneous in the domain of the regularised interfacial crack.}
	\label{fig:zoomedInterface_all}
\end{figure}

\subsection{Preliminary study}
\label{sec:prelimstudy}
First, a study has been carried out on a Double Cantilever Beam (DCB), depicted in Figure~\ref{fig:prelimdcb_geom}. The geometry 
and elastic properties are given in~\cite{irzal_isogeometric_2014}. Within the diffuse interface (red area) the fracture 
toughness is fifty times smaller than that in the surrounding bulk material: 
$\mathcal{G}_\text{c}^\text{bulk}=50\mathcal{G}_\text{c}^\text{int}$, \textit{i.e.} the fracture toughness varies abruptly. For an internal length scale~$l_\text{c} =\SI{50}{\micro\meter}$ the 
interface half width $b$ has been varied ($b = \lbrace 62.5, 31.25, 15.625, 7.8125 \rbrace\,\si{\micro\meter}$). The resulting force-displacement curves 
are given in Figure~\ref{fig:prelimdcb_fd}. A narrower interface clearly leads to a material strengthening and a higher energy dissipation.

This observation can be explained using Figure~\ref{fig:zoomedInterface_all}. In Figure~\ref{fig:zoomedInterface_all}, on the right, the phase-field for a fully developed crack has been plotted. As long as 
the crack fully lies within the interface, denoted by the solid (red) lines, the dissipation is only influenced by the fracture 
toughness of the interface, $\mathcal{G}_\text{c}^\text{int}$. When, however, the ratio between the half-width $b$ and the internal
length scale $l_\text{c}$, $b/l_\text{c}$, becomes smaller, and intersects the phase-field of a crack in the bulk material, denoted
by the dashed (red) lines, the dissipation is no longer determined by $\mathcal{G}_\text{c}^\text{int}$ alone, but is also 
influenced by $\mathcal{G}_\text{c}^\text{bulk}$. As a consequence, the results become dependent on $b$, the half width of the 
interface. 

A straightforward solution would be to increase $b/l_\text{c}$ either by increasing the interface length scale $b$ or by
lowering the phase-field length scale $l_\text{c}$. The former may not always be practical due to certain restrictions and 
topological requirements of the interface, while the latter is restricted because the phase-field length scale can be considered
as a material parameter of the bulk. Moreover, it would be computationally expensive to numerically resolve the gradients which
result from a very small value for $l_\text{c}$. \\ 
\ \\
\begin{minipage}{\textwidth}
	\centering
\begin{minipage}[c]{0.46\textwidth}
	\centering
	\captionof{table}{Specifications of the one-dimensional bar}
	\label{tab:int_1Droddata}
	\begin{tabular}{l S}
		\toprule
		$l$ & $\SI{1}{\milli\meter}$ \\
		$E$ & $\SI{210}{\giga\pascal}$ \\
		$l_\text{c}$ & $\SI{7.5}{\micro\meter}$\\
		$A$ & $\SI{1}{\milli\meter\squared}$\\
		$\mathcal{G}_\text{c}^\text{bulk}$ & $\SI{5}{\newton\per\meter}$\\
		$\mathcal{G}_\text{c}^\text{int}$ & $\SI{2.7}{\newton\per\meter}$\\
		\bottomrule
	\end{tabular}
\end{minipage}
\begin{minipage}[c]{0.5\textwidth}
	\centering
	\includegraphics[trim={0cm 0cm 0cm 0cm},clip]{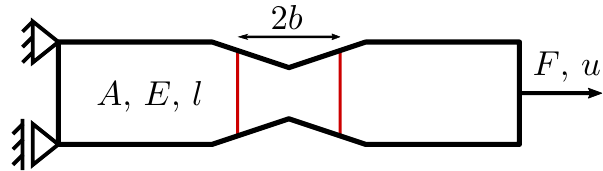}
	\captionof{figure}[1-dimensional bar geometry]{One-dimensional bar: Red lines mark bulk interface with $\mathcal{G}_\text{c}^\text{int}$ or $\hat{\mathcal{G}}_\text{c}^\text{int}$. Outside the interface, $\mathcal{G}_\text{c}^\text{bulk}$ is applied.}
	\label{fig:1D_geom}
\end{minipage}
\end{minipage}\vspace{0.5cm}

\subsection{Global dissipation equivalence}
\label{sec:int_derivation}
To quantify the influence of the bulk material on the crack propagation along the interface, the dissipated energy during crack 
growth is analysed. For this purpose, a cut perpendicular to a fully developed crack, see
Figure~\ref{fig:zoomedInterface_all}, is considered. As long as the crack phase-field does not interfere with different fracture 
toughnesses (solid red lines), the energy per unit area dissipated for a fully developed crack within the interface reads 
\begin{equation}
\label{eq:1D_correction}
D=\frac{\mathcal{G}_\text{c}^\text{int}}{4l_\text{c}}\int\limits_{-b}^{b} \left(1-c\right)^2+4l_\text{c}^2 c_{,x}^2\,\text{d}x\point
\end{equation}
The integral over the bulk material does not contribute to $D$ because the crack phase-field takes a constant value of $c\equiv1$. 
In other words, the ratio $b/l_\text{c}$ is sufficiently large: $b/l_\text{c}\rightarrow \infty$. Substituting the analytical 
description of the crack phase-field, $c=1-\exp(-\vert x \vert/(2l_\text{c}))$, into Equation~\eqref{eq:1D_correction} and 
simplifying yields:
\begin{equation}
\label{eq:1D_correction_solution}
D=\mathcal{G}_\text{c}^\text{int}\point
\end{equation}
Since the interface is not infinitely wide for realistic applications, the bulk influence has to be accounted for (red dashed lines) 
and Equation~\eqref{eq:1D_correction} is replaced by:
\begin{equation}
\label{eq:1D_correction_mod}
\hat{D}=\frac{1}{2l_\text{c}}\left[  \hat{\mathcal{G}}_\text{c}^\text{int}\int\limits_{0}^{b} \left(1-c\right)^2+4l_\text{c}^2 c_{,x}^2 \,\text{d}x+\mathcal{G}_\text{c}^\text{bulk}\int\limits_{b}^{\infty} \left(1-c\right)^2+4l_\text{c}^2 c_{,x}^2 \,\text{d}x  \right]
\end{equation} 
with $\hat{\mathcal{G}}_\text{c}^\text{int}$ the modified fracture toughness of the interface, which is adjusted to account for the 
bulk influence. Evidently, the second term in brackets needs not to be taken into account in Equation~\eqref{eq:1D_correction} since $b$ is then sufficiently large, $b/l_\text{c}\rightarrow\infty$. Elaboration of Equation~\eqref{eq:1D_correction_mod} yields:
\begin{equation}
\label{eq:1D_correction_mod_solution}
\hat{D}= \hat{\mathcal{G}}_\text{c}^\text{int} \left( 1-\text{e}^{-b/l_\text{c}} \right) + \mathcal{G}_\text{c}^\text{bulk}\,\text{e}^{-b/l_\text{c}}\point
\end{equation}
We now require that $D=\hat{D}$. Rearranging Equations~\eqref{eq:1D_correction_solution} and \eqref{eq:1D_correction_mod_solution} 
gives the modified interface fracture toughness:
\begin{equation}
\label{eq:gc_corr}
\boxed{ \hat{\mathcal{G}}_\text{c}^\text{int}=\frac{\mathcal{G}_\text{c}^\text{int}-\mathcal{G}_\text{c}^\text{bulk}\,\text{e}^{-b/l_\text{c}}}{1-\text{e}^{-b/l_\text{c}}} } \quad\text{with}\quad\frac{b}{l_\text{c}}>\ln\frac{\mathcal{G}_\text{c}^\text{bulk}}{\mathcal{G}_\text{c}^\text{int}}
\end{equation}
where the latter constraint ensures that $\hat{\mathcal{G}}_\text{c}^\text{int}$ does not take unphysical values, i.e. becomes 
smaller than zero. This approach would also work for an inverse correction, i.e. when 
$\mathcal{G}_\text{c}^\text{bulk} < \mathcal{G}_\text{c}^\text{int}$. Herein we have limited ourselves, however, to cases where 
$\mathcal{G}_\text{c}^\text{bulk} > \mathcal{G}_\text{c}^\text{int}$.

\subsection{One-dimensional bar}
\label{sec:int_1D}
\begin{table}[b]
	\centering
	\caption{Dissipated energy for one-dimensional bar: For the setup where the original interface fracture toughness is used, the dissipated energy depends on the ratio $b/l_\text{c}$ for small values. Using the compensated, artificially lowered interface fracture toughness the dissipated energy is almost equal for each case. The homogeneous case where $b/l_\text{c}\rightarrow\infty$ serves as reference $\Psi^\text{c,ref}$.}
	\label{tab:1D_energy}
	\begin{tabular}{l l| c c  c c c }
		\toprule
		\multicolumn{2}{r}{$b/l_\text{c}$}  & $\rightarrow\infty$  &  8.33 & 4.17 & 2.08 \\
		\multirow{2}{2.5cm}{$\Psi^\text{c}/\Psi^\text{c,ref}-1$~using}& $\mathcal{G}_\text{c}^\text{int}$ &   0 & 0 & 0.013 & 0.109 \\
		& $\hat{\mathcal{G}}_\text{c}^\text{int}$ &   0 & 0 & -0.01 & -0.011\\
		\bottomrule
	\end{tabular}
\end{table}
One-dimensional numerical studies have been carried out to demonstrate the effect of the correction on the force-displacement 
curves. The test case is a one-dimensional bar \mbox{$x \in [0,1]\,\si{\milli\meter}$} with a reduced cross-sectional area in the
centre to nucleate a crack, see Figure~\ref{fig:1D_geom}, while the material properties are given in 
Table~\ref{tab:int_1Droddata}. The reduced cross section has been accounted for by introducing a variable cross-sectional area in 
the derivation. The simulations have been carried out with a one-dimensional phase-field code which uses hierarchical B-spline 
basis functions. Quadratic shape functions have been used with reduced $C^0$ continuity at the centre of the bar. The spatial 
discretisation can be considered as sufficiently fine~\cite{linse_convergence_2017}. Now, the dissipated energy $\Psi^\text{c}$ is compared to a reference value $\Psi^\text{c,ref}$ obtained from a 1-dimensional bar with $\mathcal{G}_\text{c}^\text{int}$. It is expected, that the dissipated energy is highly overestimated without the compensation, which is indeed the case:  A comparison between the cases with and
without the correction is shown in Table~\ref{tab:1D_energy} with respect to the dissipated energy $\Psi^\text{c}$ for the fully 
cracked bar. The correction clearly has the intended effect and brings down the deviation of more than 10\% for the smallest ratio presented.

\section{Phase-field Modelling of Interface Failure: Extension to two dimensions}
\label{sec:interface2D}

\begin{figure}[tb]
	\centering
	\subfloat[Geometry and boundary conditions ]{
		\label{fig:geom_BC}
		\includegraphics[trim={0cm 0cm 0cm 0cm},clip]{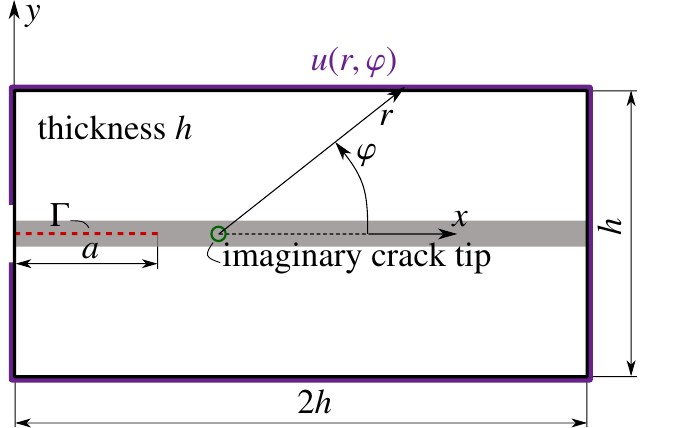}
	}%\hspace{1cm}
	\subfloat[Enlarged initial notch]{
		\label{fig:geom_iniPF}
		\raisebox{0.26cm}{\includegraphics[trim={0.85cm 0.5cm 0cm 0.5cm},clip]{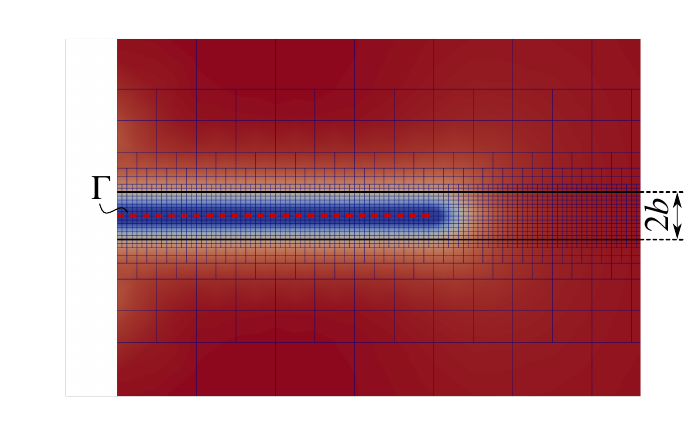}}
	}\\
	\subfloat[Undeformed geometry]{
		\label{fig:undef_mesh}
		\includegraphics[trim={0cm 0cm 9cm 0cm},clip]{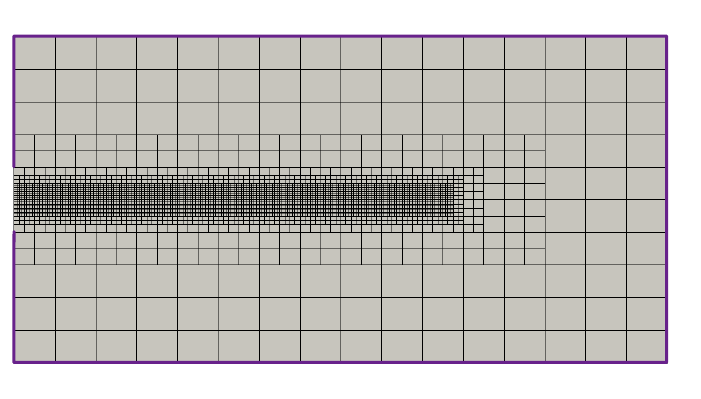}
	} 
	\subfloat[Deformed geometry]{
		\label{fig:def_mesh}
		\includegraphics[trim={9cm 0cm 0cm 0cm},clip]{fig5cd.pdf}
	}
	\caption[Geometry and boundary conditions]{Geometry and boundary conditions: (a) The geometry with $h=2a=\SI{1}{\milli\meter}$ is described. During the simulation, an imaginary crack tip (green circle) propagates along the $x$-axis and the near field K-concept displacements $u(r,\varphi)$ of a mode-I crack are assigned to all nodes on the purple edges. The grey stripe resembles the diffuse interface. (b) The predefined notch is initialised by setting the phase-field $c=0$ (blue) along the red dashed line. The diffuse interface of width $2b$ is varied. In (c) and (d) the undeformed and deformed geometries are depicted. The displacements in (d) are exaggerated.}
	\label{fig:geom_BC_iniPF}
\end{figure}

\subsection{Two-dimensional double cantilever beam}
The testing geometry in two dimensions is a modified Double Cantilever Beam (DCB) with an initial notch, see 
Figure~\ref{fig:geom_BC}. It differs from a classical DCB in the sense that the specimen is deformed using a so-called 
`surfing boundary condition'~\cite{kuhn_discussion_2016,hossain_effective_2014}, which projects the near-field displacements 
$u(r,\varphi)$ of a mode-I crack obtained in linear elastic fracture mechanics to the IGA control points of the boundary edges. In this sense, 
an imaginary crack tip propagates along the $x$-axis and tears apart the upper and lower halves of the DCB. 
Figures~\ref{fig:undef_mesh}~and~\ref{fig:def_mesh} give the undeformed and the deformed meshes, respectively. In the 
two-dimensional simulations, there is not only crack initiation, but also crack propagation, and therefore the energy release rate 
is evaluated as well, and is compared with the fracture toughness. As will be demonstrated below, the surfing boundary condition 
yields steady crack growth and enables the accurate determination of the energy release rate. The diffuse interface, i.e. the grey 
shaded area in Figure~\ref{fig:geom_BC}, is aligned with the $x$-axis and is incorporated by assigning different values $\mathcal{G}_\text{c}^\text{bulk}$ and $\mathcal{G}_\text{c}^\text{int}$ for the 
fracture toughness outside and inside the interface, \textit{i.e.} the fracture toughness varies abruptly.
The mesh is refined along the expected crack path. The initial crack is incorporated by setting the phase-field $c=0$ (blue) along 
the straight notch, see Figure~\ref{fig:geom_iniPF}. The averaged \mbox{$y$-displacement} $\bar{u}$ of the nodes along the upper 
edge serves as abscissa for the following plots. The following values have been used for the material parameters: 
\textsc{Young}'s modulus $E=\SI{210}{\giga\pascal}$, \textsc{Poisson}'s ratio $\nu=0.3$ and $l_\text{c}=\SI{15}{\micro\meter}$.
Plane strain conditions are assumed. The values for the fracture toughness are set to $\mathcal{G}_\text{c}^\text{bulk}=\SI{5.4}{\newton\per\milli\meter}$ and 
$\mathcal{G}_\text{c}^\text{int}=\SI{2.7}{\newton\per\milli\meter}$. The interface half-width $b$ has been varied and the 
'compensated' interface fracture toughness $\hat{\mathcal{G}}_\text{c}^\text{int}$ has been calculated according to 
Equation~\eqref{eq:gc_corr}.  

The energy release rate $\mathcal{G}$ has been calculated using the concept of configurational forces~\cite{kuhn_discussion_2016},
which is comparable to a generalised $J$-integral~\cite{kuna_numerische_2008}. It is important to note that the predefined fracture 
toughness $\mathcal{G}_\text{c}$ is not recovered numerically. Indeed, the discretisation and the choice of the length scale alter 
the predefined value~\cite{bourdin_variational_2008} and the numerical fracture toughness
\begin{equation}
\label{eq:gc_num}
\mathcal{G}_\text{c,num}=\mathcal{G}_\text{c}\cdot\left(1+\frac{\Delta h}{4 l_\text{c}}\right)
\end{equation}
governs the simulation. Herein, $\Delta h=\SI{6.25}{\micro\meter}$ is the characteristic mesh spacing.

\begin{figure}[t]
	\centering
	\subfloat[Original curves  ]{
		\label{fig:2D_CE_nC}
		\includegraphics[trim={0cm 0cm 0cm 0cm},clip]{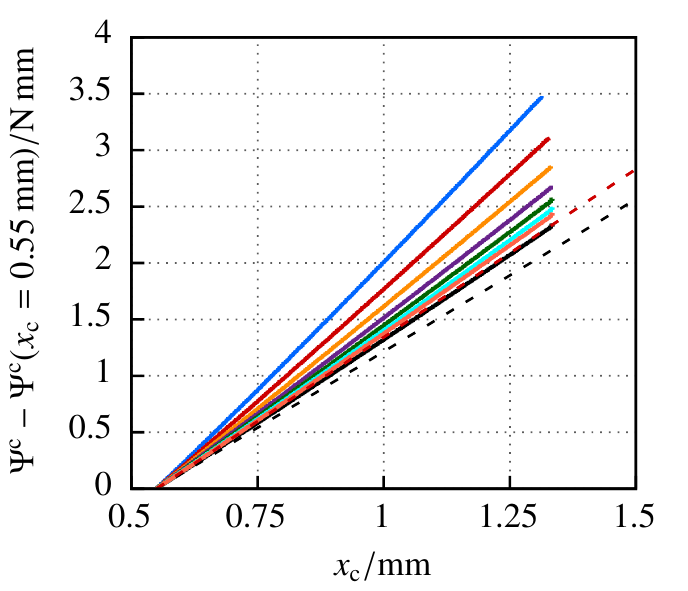}
	}
	\subfloat[Compensated curves ]{
		\label{fig:2D_CE_C}
		\includegraphics[trim={0cm 0cm 0cm 0cm},clip]{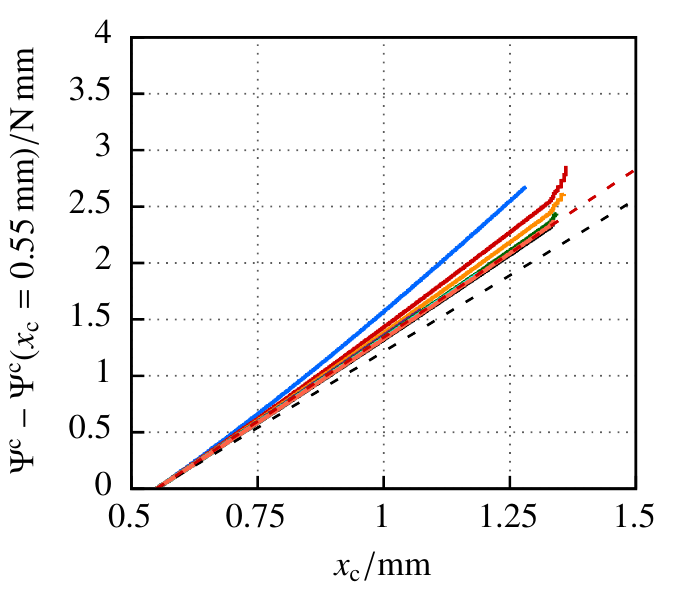}
	}
	\\
	\vspace{0.3cm}
	\footnotesize 
	\begin{tabular}{l|l|l|l||l}
		\legline{1} Reference & \legline{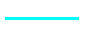} $b/l_\text{c}=2.92$ & \legline{5} $b/l_\text{c}=2.08$ & \legline{3} $b/l_\text{c}=1.25$ & \legline{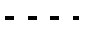} $\mathcal{G}_\text{c}^\text{int}\cdot\Delta x_\text{c}$\\    
		\legline{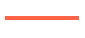} $b/l_\text{c}=3.33$ & \legline{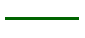} $b/l_\text{c}=2.5$ & \legline{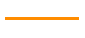} $b/l_\text{c}=1.67$ & \legline{2} $b/l_\text{c}=0.83$ & \legline{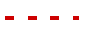} $\mathcal{G}_\text{c,num}^\text{int}\cdot\Delta x_\text{c}$\\
	\end{tabular}	
	
	\caption[Global dissipation equivalence -- crack energy]{Crack energy: (a) Before the correction, the crack energy increases faster for smaller interface widths. (b) Applying the correction balances the crack energy increase to the same level. For small interface widths (red/blue curve) deviations occur. A closer look reveals that these curves are not linear, which implies that the crack length is overestimated.  }
	\label{fig:2D_CE}
\end{figure}

Figure~\ref{fig:2D_CE} shows the increase of the dissipated energy from the point where the crack tip passes 
$x=x_\text{c}=\SI{0.55}{\milli\meter}$. The value has been chosen slightly higher than the initial crack length to rule out 
differences due to the crack initialisation. The actual crack energy increase is compared with the theoretical value, i.e. the 
fracture toughness multiplied by the finite crack growth $\Delta x_\text{c}$. It is noted that the crack is an actual surface 
with $h=\SI{1}{\milli\meter}$ in the out-of-plane direction. As expected, the reference with $b/l_\text{c}\rightarrow\infty$ 
coincides with the theoretical increase represented by the (red) dashed line. For smaller ratios $b/l_\text{c}$ similar to the case of the one-dimensional bar, the 
dissipated energy is overestimated due to the bulk material influence, which is evidenced from the increasing inclination of the 
curves in Figure~\ref{fig:2D_CE_nC}. For ratios up to $b/l_\text{c}\approx2$, the correction suggested in the preceding section works 
fairly well, see Figure~\ref{fig:2D_CE_C}. But for small ratios of $b/l_\text{c}$ an increasing discrepancy occurs. One reason can be the discretisation of the phase-field which is considerably coarser compared to the one-dimensional case. Apart from this, the global compensation approach also yields a significant deviation for the critical energy release rate as will be pointed out in the following.

\begin{figure} 
	\centering
	\subfloat[$\mathcal{G}$ and crack growth]{
		\label{fig:pre_ERR}
		\includegraphics[trim={0cm 0cm 0cm 0cm},clip]{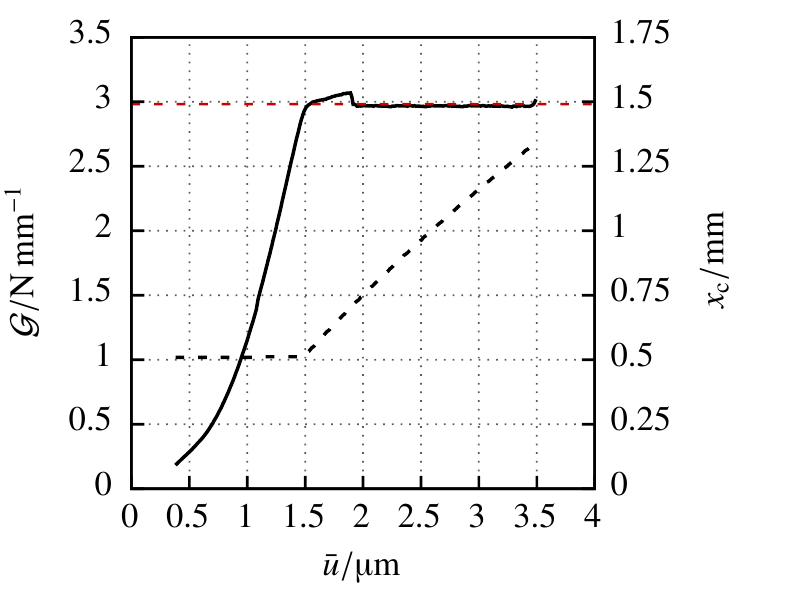}
	}
	\subfloat[Bulk influence on $\mathcal{G}_\text{c}$ -- original curves ]{
		\label{fig:pre_bulkinf}
		\includegraphics[trim={0cm 0.5cm 1cm 1.5cm},clip]{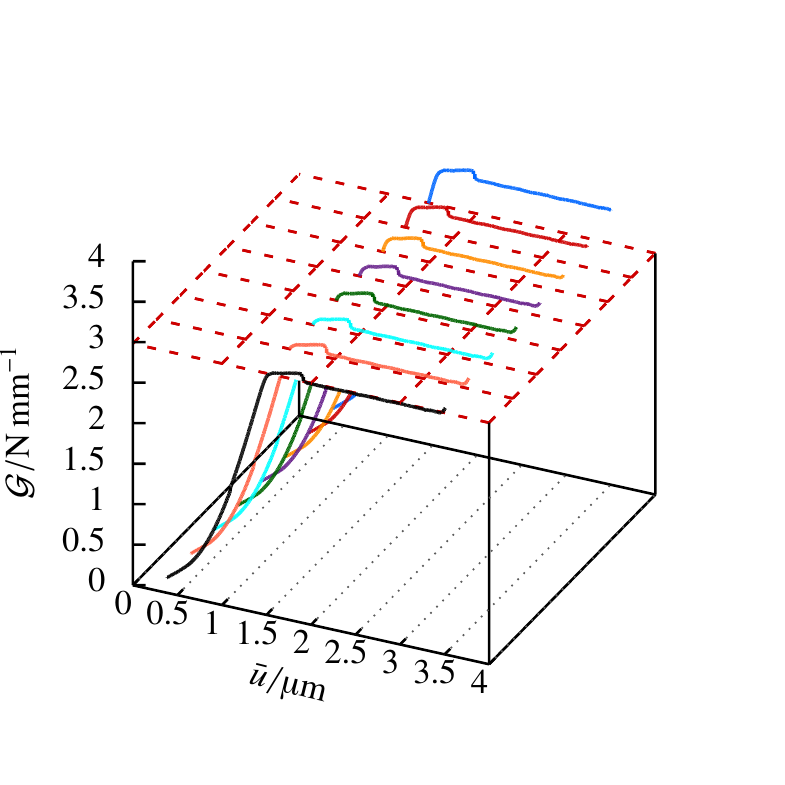}
	}\\
	\subfloat[Bulk influence on $\mathcal{G}_\text{c}$ -- compensation from Sec.~\ref{sec:int_derivation} applied ]{
		\label{fig:pre_bulkinf_compen}
		\includegraphics[trim={0cm 0cm 0cm 0cm},clip]{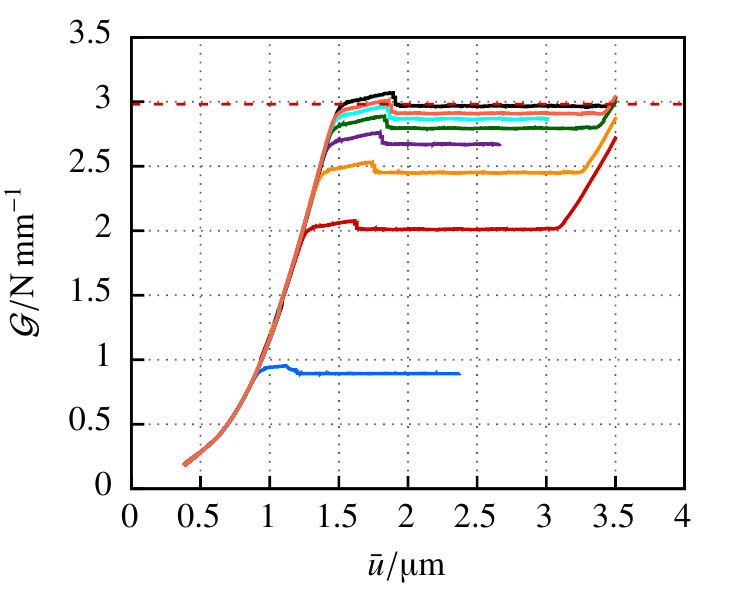}
	}\\
	\vspace{0.3cm}
	\footnotesize 
	\begin{tabular}{l|l|l|l||l}
		\legline{1} Reference & \legline{7} $b/l_\text{c}=2.92$ & \legline{5} $b/l_\text{c}=2.08$ & \legline{3} $b/l_\text{c}=1.25$ & \legline{10} Reference $x_\text{c}$\\    
		\legline{8} $b/l_\text{c}=3.33$ & \legline{6} $b/l_\text{c}=2.5$ & \legline{4} $b/l_\text{c}=1.67$ & \legline{2} $b/l_\text{c}=0.83$ & \legline{12} $\mathcal{G}_\text{c,num}^\text{int}$\\
	\end{tabular}	
	
	\caption{$\mathcal{G}$ and crack length: (a) The energy release rate $\mathcal{G}$ and the crack tip position $x_\text{c}$ are plotted as a function of the average displacement $\bar{u}$. As soon as $\mathcal{G}$ reaches the numerical fracture toughness $\mathcal{G}_\text{c,num}^\text{int}$ the crack starts to propagate. The bumps in the $\mathcal{G}$ curves are artefacts due to the phase-field initialisation. (b) The fracture toughness governing crack growth is between the bulk and the interface fracture toughness and increases for smaller ratios $b/l_\text{c}$ because the bulk influence becomes more significant. It is noted, that the 3-dimensional plot layout has only been chosen for visualisation purposes because certain curves would not be differentiable in a 2-dimensional plot. (c) The global dissipation equivalence approach results in a large underestimation of the energy release rate of the crack propagating along the interface.}
	\label{fig:pre_ERR_bulkinf}
\end{figure}

In the reference case, i.e. $b/l_\text{c}\rightarrow\infty$, the entire domain is assigned the interface fracture 
toughness~$\mathcal{G}_\text{c}^\text{int}=\SI{2.7}{\newton\per\milli\meter}$ and the bulk fracture toughness does not play a role.
In contrast to a one-dimensional setting, the crack propagates for a two-dimensional situation, i.e. the $x$-position of the crack tip~$x_\text{c}$ changes when $\mathcal{G}$ has reached a critical 
value, see Figure~\ref{fig:pre_ERR}. Again we observe an interaction between the values of the fracture toughness for the bulk material and for the regularised interface. For further illustration, the ratio $b/l_\text{c}$ has been reduced gradually starting from the reference case. The resulting energy release rates are compared in 
Figure~\ref{fig:pre_bulkinf}. It is clear that $\mathcal{G}_\text{c}$ attains higher values for smaller ratios $b/l_\text{c}$,
which is consistent with the results reviewed in Sections~\ref{sec:prelimstudy} and \ref{sec:int_1D}. 

Directly applying the compensation approach of Sec.~\ref{sec:int_derivation} yields a significant underestimation of the energy release rate for crack 
propagation as is evident from Figure~\ref{fig:pre_bulkinf_compen}. This is a drawback of the compensation method as it has
been formulated for the one-dimensional case. Indeed, for $b/l_\text{c}\rightarrow0$ the method can no longer be applied since 
the subsidiary condition formulated in Equation~(\ref{eq:gc_corr}) is no longer fulfilled. A closer look at the compensation 
process illustrates the problem. The requirement of a global dissipation equivalence takes into account also areas where the crack 
has already propagated and is zero for intact parts of the body. However, the actual crack growth is governed by physical 
quantities in a region of the order $l_\text{c}$ in front of the crack tip. 

All in all, the global dissipation equivalence approach does not account for crack propagation, as it is the case for two-dimensional simulations, and is only valid for a one-dimensional setup where a crack emerges. However, it is essential to consider the energy release rate in order to predict crack growth. This raises the need for another compensation method. 
\subsection{Analytical implications using the concept of configurational forces}
Kuhn and Müller~\cite{kuhn_discussion_2016} have derived a quantitative crack growth criterion, which accounts for heterogeneities in the vicinity of the crack tip. They showed that 
the $x$-component of the configurational force which contains the phase-field contributions 
$\boldsymbol{\mathcal{G}}^\text{fr}$ is equal to the fracture toughness when the crack starts propagating.  It is noted, that the other contributions to the configurational force listed in~\cite{kuhn_discussion_2016} are zero for crack propagation within this contribution.
The configurational force can be written as an integral over a finite area $A$ around the crack tip of the divergence of the 
\textsc{Eshelby} tensor:
\begin{equation}
\label{eq:gfr}
\mathcal{G}^\text{fr}_j=-\int\limits_A \Sigma_{ij,i}^\text{fr} \,\text{d}A\,\coma
\end{equation}
where
\begin{equation}
\Sigma_{ij}^\text{fr}=\psi^\text{c}\delta_{ij}-2\mathcal{G}_\text{c}l_\text{c}c_{,i}c_{,j}\point
\end{equation}
Making use of the analytical one-dimensional solution of the phase-field $c$, the integral over $A$ can be split into two contributions $A=A_\text{crack}+A_\text{tip}$ and 
evaluated afterwards. For a crack propagating straight along the $x$-axis, the first contribution $A_\text{crack}$ stems from the 
fully developed straight crack to the left of the crack tip. Here, the phase-field is homogeneous in the $x$-direction and follows 
the one-dimensional solution in the $y$-direction. Evaluating the $x$-component of the integral for $A_\text{crack}$ in 
Equation~\eqref{eq:gfr} yields $\mathcal{G}^\text{fr}_1\equiv0$, i.e. crack growth is not influenced by the tip history. 
Consequently, only the phase-field around the crack tip within $A_\text{tip}$ contributes to $\mathcal{G}^\text{fr}_1$. The analytical evaluation of this 
contribution is challenging since the circumferential phase-field solution at the crack tip, which is attached to the straight 
crack path, is not known a priori. This also applies to the analytical solution of the two-dimensional \textsc{Helmholtz} 
differential equation for this setup. For these reasons, a numerical approach is chosen below to quantify the influence of the 
fracture toughness of the  interface and the bulk material, thus avoiding an analytical evaluation of Equation~\eqref{eq:gfr}.

\subsection{Numerical study for a local compensation approach}
\begin{figure}[t]
	\centering
	\subfloat[Generalised exaggeration ]{
		\label{fig:lca_deriv}
		\includegraphics[trim={0cm 0cm 0cm 0cm},clip]{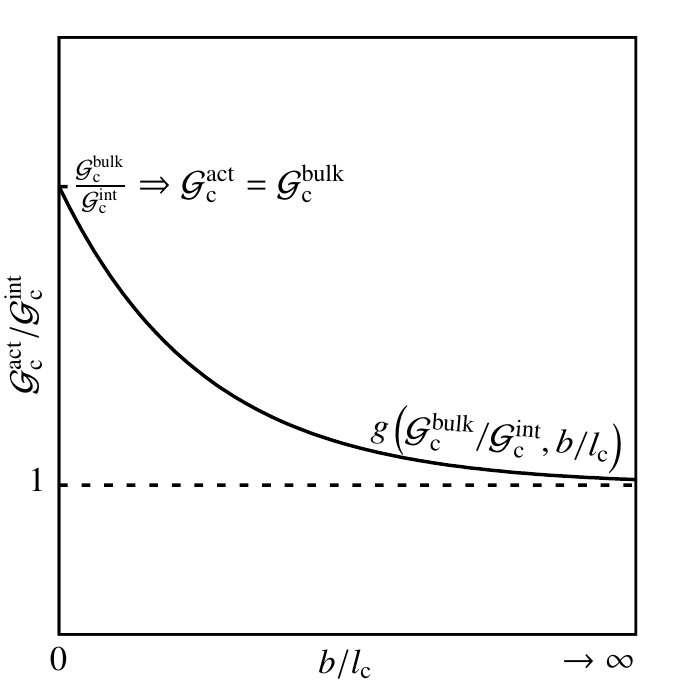}
	}
	\subfloat[Simulation data]{
		\label{fig:lca_sim}
		\includegraphics[trim={0cm 0cm 0cm 0cm},clip]{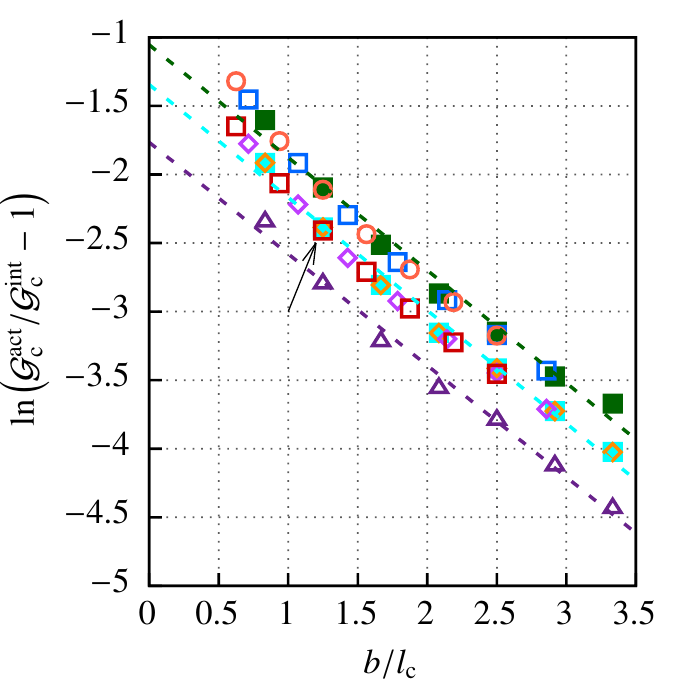}
	}
	\\
	\vspace{0.3cm}
	\footnotesize 
	\begin{tabular}{l l||l }
		\multirow{ 4}{*}{\hspace{0.2cm}$l_\text{c}=\SI{15}{\micro\meter} \left\{ \begin{array}{c}
			\vspace{0.1cm} \\
			\\
			\\ 
			\\
			\end{array} \right.$}\hspace{-0.5cm}		& \legpoint{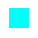} $\mathcal{G}_\text{c}^\text{bulk}/\mathcal{G}_\text{c}^\text{int}=2$, $\mathcal{G}_\text{c}^\text{bulk}=\SI{5.4}{\newton\per\milli\meter}$, $\mathcal{G}_\text{c}^\text{int}=\SI{2.7}{\newton\per\milli\meter}$&  \multirow{ 3}{*}{$ \left. \begin{array}{c}
			\text{\legline{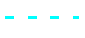}} \\
			\text{\legline{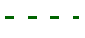}} \\
			\text{\legline{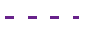}}			
			\end{array} \right\}$ exponential fits for $g$}\\    
		& \legpoint{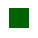} $\mathcal{G}_\text{c}^\text{bulk}/\mathcal{G}_\text{c}^\text{int}=3$, $\mathcal{G}_\text{c}^\text{bulk}=\SI{8.1}{\newton\per\milli\meter}$, $\mathcal{G}_\text{c}^\text{int}=\SI{2.7}{\newton\per\milli\meter}$ & \\
		& \legpoint{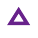} $\mathcal{G}_\text{c}^\text{bulk}/\mathcal{G}_\text{c}^\text{int}=1.5$, $\mathcal{G}_\text{c}^\text{bulk}=\SI{4.05}{\newton\per\milli\meter}$, $\mathcal{G}_\text{c}^\text{int}=\SI{2.7}{\newton\per\milli\meter}$& \\
		& \legpoint{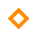} $\mathcal{G}_\text{c}^\text{bulk}/\mathcal{G}_\text{c}^\text{int}=2$, $\mathcal{G}_\text{c}^\text{bulk}=\SI{8.1}{\newton\per\milli\meter}$, $\mathcal{G}_\text{c}^\text{int}=\SI{4.05}{\newton\per\milli\meter}$ &\\
		\multirow{ 2}{*}{$l_\text{c}=\SI{17.5}{\micro\meter} \left\{ \begin{array}{c}
			\\
			\\
			\end{array} \right.$}\hspace{-0.5cm} & \legpoint{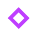} $\mathcal{G}_\text{c}^\text{bulk}/ \mathcal{G}_\text{c}^\text{int}=2$, $\mathcal{G}_\text{c}^\text{bulk}= \SI{5.4}{\newton\per\milli\meter}$, $\mathcal{G}_\text{c}^\text{int}= \SI{2.7}{\newton\per\milli\meter}$ &  \\
		&\legpoint{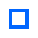} $\mathcal{G}_\text{c}^\text{bulk}/ \mathcal{G}_\text{c}^\text{int}=3$, $\mathcal{G}_\text{c}^\text{bulk}= \SI{8.1}{\newton\per\milli\meter}$, $\mathcal{G}_\text{c}^\text{int}= \SI{2.7}{\newton\per\milli\meter}$ & \\
		\multirow{ 2}{*}{\hspace{0.2cm}$l_\text{c}=\SI{20}{\micro\meter} \left\{ \begin{array}{c}
			\\
			\\
			\end{array} \right.$}\hspace{-0.5cm}	& \legpoint{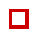} $\mathcal{G}_\text{c}^\text{bulk}/\mathcal{G}_\text{c}^\text{int}=2$, $\mathcal{G}_\text{c}^\text{bulk}=\SI{5.4}{\newton\per\milli\meter}$, $\mathcal{G}_\text{c}^\text{int}=\SI{2.7}{\newton\per\milli\meter}$ & \\
		& \legpoint{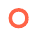} $\mathcal{G}_\text{c}^\text{bulk}/\mathcal{G}_\text{c}^\text{int}=3$, $\mathcal{G}_\text{c}^\text{bulk}=\SI{8.1}{\newton\per\milli\meter}$, $\mathcal{G}_\text{c}^\text{int}=\SI{2.7}{\newton\per\milli\meter}$ & \\
	\end{tabular}	
	
	\caption{Schematic illustration and numerical examples: (a) The sensitivity of the actual fracture toughness $\mathcal{G}_\text{c}^\text{act}$, which the material experiences, to the ratio $b/l_\text{c}$ is shown. The exaggeration function $g$ which describes the relationship is unknown, but an exponential relation is likely, since the phase-field follows such a description. (b) Simulation data reinforcing the use of $g$ are presented. Over a wide range $b/l_\text{c}$ the data behave linearly, independent from the absolute values of $b$ and $l_\text{c}$. Importantly, the exponential fits show that the curve grouping only depends on the ratio $\mathcal{G}_\text{c}^\text{bulk}/\mathcal{G}_\text{c}^\text{int}$.}
	\label{fig:lca_deriv_sim}
\end{figure}

Figure~\ref{fig:lca_deriv} postulates a general relation between the ratios $b/l_\text{c}$, 
$\mathcal{G}_\text{c}^\text{bulk}/\mathcal{G}_\text{c}^\text{int}$ and the fracture toughness, which is numerically measured during the simulations. This fracture toughness is referred to as actual fracture toughness 
$\mathcal{G}_\text{c}^\text{act}$. The above observations suggest, that the ratio
\begin{equation}
\frac{\mathcal{G}_\text{c}^\text{act}}{\mathcal{G}_\text{c}^\text{int}}=g\left( \frac{\mathcal{G}_\text{c}^\text{bulk}}{\mathcal{G}_\text{c}^\text{int}},\frac{b}{l_\text{c}} \right)
\end{equation}
is a priori unknown. The function $g$ will be referred to as the \textit{exaggeration} function, \textit{i.e.} the interface fracture toughness is exaggerated because of the bulk material influence and, for the most general case, one ends up with an actual fracture toughness not equal to the interface fracture toughness.

Now, two limiting cases will be discussed. For $b/l_\text{c} \rightarrow \infty$ the actual, numerically measured fracture 
toughness $\mathcal{G}_\text{c,num}^\text{act}$ equals the numerical interface fracture toughness 
$\mathcal{G}_\text{c,num}^\text{int}$:
\begin{equation}
\frac{\mathcal{G}_\text{c,num}^\text{act}}{\mathcal{G}_\text{c,num}^\text{int}}=
\frac{\mathcal{G}_\text{c}^\text{act}}{\mathcal{G}_\text{c}^\text{int}}=1 \point
\end{equation}
This case is similar to the example in Figure~\ref{fig:pre_ERR}. Please note that the discretisation effect in 
Equation~\eqref{eq:gc_num} cancels when considering relative values of the exaggeration function and is only relevant when 
interpreting absolute numerical values. For $b/l_\text{c} \rightarrow 0$ the actual, numerically measured value of the fracture 
toughness $\mathcal{G}_\text{c}^\text{act}$ equals the fracture toughness of the bulk material $\mathcal{G}_\text{c}^\text{bulk}$:
\begin{equation}
\frac{\mathcal{G}_\text{c}^\text{act}}{\mathcal{G}_\text{c}^\text{int}}=\frac{\mathcal{G}_\text{c}^\text{bulk}}{\mathcal{G}_\text{c}^\text{int}}\point
\end{equation} 
The range of the ratios $\mathcal{G}_\text{c}^\text{bulk}/\mathcal{G}_\text{c}^\text{int}$ has been chosen such that it can be
compared with analytical results from linear elastic fracture mechanics~\cite{he_crack_1989}, 
cf. Figure~\ref{fig:angle_deflection}. 

Figure~\ref{fig:lca_sim} presents the semi-logarithmic plots for a large variety of different material parameters. Despite 
significant differences in the length scales of the regularisations and the corresponding values of the fracture toughnesses, 
a pattern can be observed for the non-dimensional results. All the results with the same ratio 
$\mathcal{G}_\text{c}^\text{bulk}/\mathcal{G}_\text{c}^\text{int}$ form a group independent from the ratio $b/l_\text{c}$. This 
is emphasised by the three fits. The exaggeration function only depends on ratios of the values of the fracture toughnesses and the
regularisation length scales, but not on the absolute values. The arrow in Figure~\ref{fig:lca_sim} points at a representative 
example: The blue filled square, orange diamond and red non-filled square stand for simulations with different absolute values of 
the fracture toughnesses and the length scale $l_\text{c}$, but the value of the exaggeration function is the same for each 
simulation. This points at a general relationship for the exaggeration function $g$. Motivated by the exponential shape of 
the phase-field perpendicular to the crack path and the one-dimensional findings, the exaggeration function is chosen to be of an exponential type:
\begin{equation}
\label{eq:int_alternativeCorrFunc}
\frac{\mathcal{G}_\text{c}^\text{act}}{\mathcal{G}_\text{c}^\text{int}}= g\left(\frac{\mathcal{G}_\text{c}^\text{bulk}}{\mathcal{G}_\text{c}^\text{int}},\frac{b}{l_\text{c}}\right) = Q \exp\left(-r \cdot\frac{b}{l_\text{c}} \right) + 1\coma
\end{equation}
where the constants $Q$ and $r$ may depend on the ratios $b/l_\text{c}$ and 
$\mathcal{G}_\text{c}^\text{bulk}/\mathcal{G}_\text{c}^\text{int}$. 

\begin{figure}[p]
	\centering
	\subfloat[Curve family for compensation factor]{
		\label{fig:compensation_large}
		\includegraphics[trim={0cm 0cm 0cm 0cm},clip]{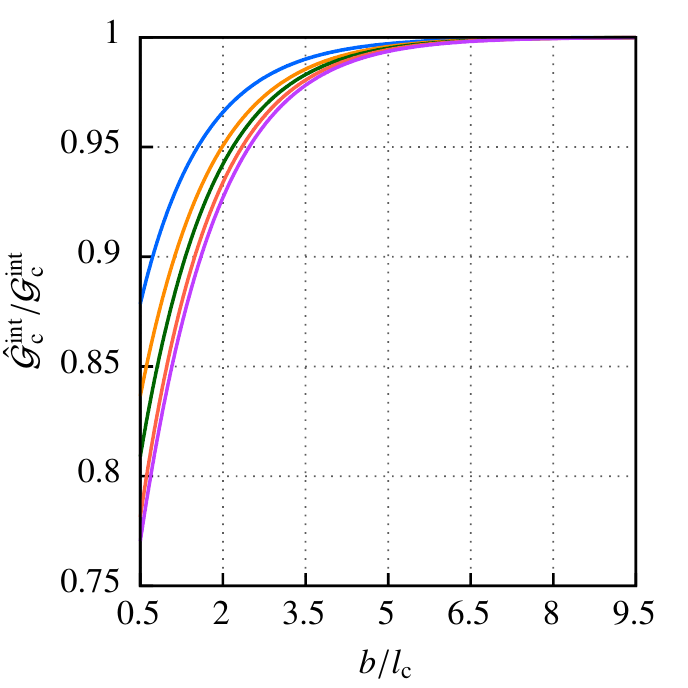}
	}
	\subfloat[Two examples for bulk influence compensation]{
		\label{fig:compensation_small}
		\includegraphics[trim={0cm 0cm 0cm 0cm},clip]{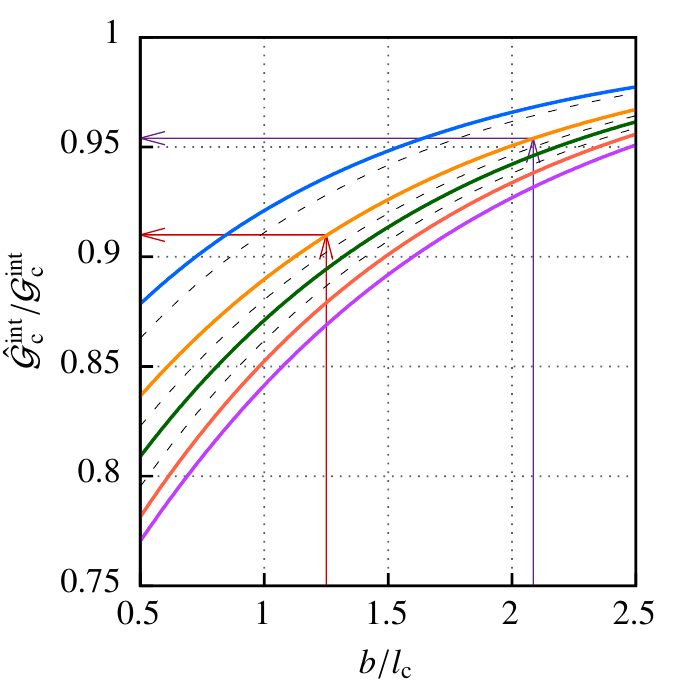}
	}\\
	\vspace{0.3cm}
	\footnotesize 
	\begin{tabular}{l|l|l|l|l}
		\legline{2} $\mathcal{G}_\text{c}^\text{bulk}/\mathcal{G}_\text{c}^\text{int}\equiv 1.5$ & \legline{4} $\mathcal{G}_\text{c}^\text{bulk}/\mathcal{G}_\text{c}^\text{int}\equiv 2$ & \legline{6} $\mathcal{G}_\text{c}^\text{bulk}/\mathcal{G}_\text{c}^\text{int}\equiv 2.5$ &  \legline{8} $\mathcal{G}_\text{c}^\text{bulk}/\mathcal{G}_\text{c}^\text{int}\equiv 3$ &\legline{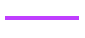} $\mathcal{G}_\text{c}^\text{bulk}/\mathcal{G}_\text{c}^\text{int}\equiv 4$\\    
	\end{tabular}	
	
	\caption{Compensation of bulk influence: (a) Each curve is valid for a constant ratio $\mathcal{G}_\text{c}^\text{bulk}/\mathcal{G}_\text{c}^\text{int}$. For finite ratios $b/l_\text{c}$ the bulk influence has to be compensated by artificially lowering the original interface fracture toughness $\mathcal{G}_\text{c}^\text{int}$ by a certain factor depending on $b/l_\text{c}$. The factor can be determined from the ordinate. As mentioned, the bulk influence vanishes for large ratios $b/l_\text{c}$ which can be seen from the curves converging to $\hat{\mathcal{G}}_\text{c}^\text{int}/\mathcal{G}_\text{c}^\text{int}=1$. (b) For two examples with a ratio $\mathcal{G}_\text{c}^\text{bulk}/\mathcal{G}_\text{c}^\text{int}\equiv 2$ the compensation factor is determined. The thin dashed black lines indicate more curves belonging to the same family.}
	\label{fig:compensation}
	\vspace{0.5cm}
	\subfloat[Five curve families with $b/l_\text{c}\equiv\text{const}$ ]{
	\includegraphics{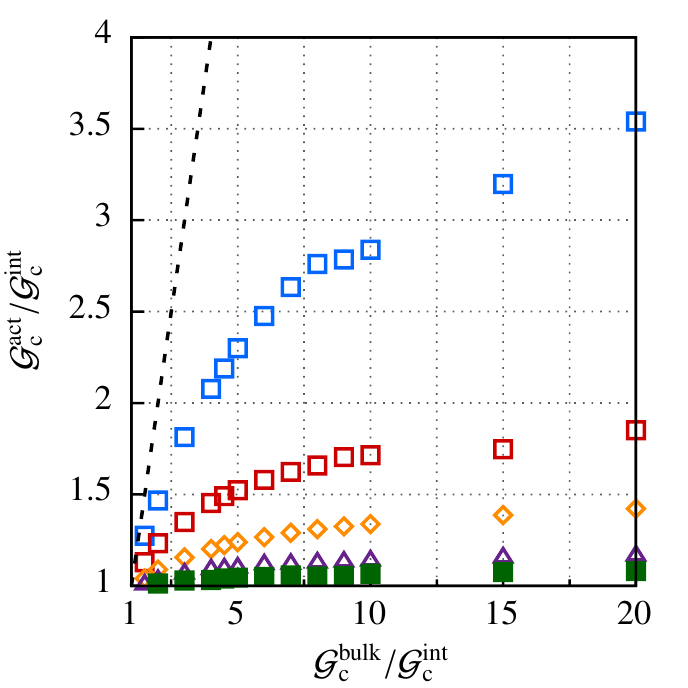}
	\label{fig:largerat}}
	\subfloat[Curve families with $\mathcal{G}_\text{c}^\text{bulk}/\mathcal{G}_\text{c}^\text{int}\equiv\text{const}$ analogue to Fig.~\ref{fig:lca_sim}  ]{
		\includegraphics{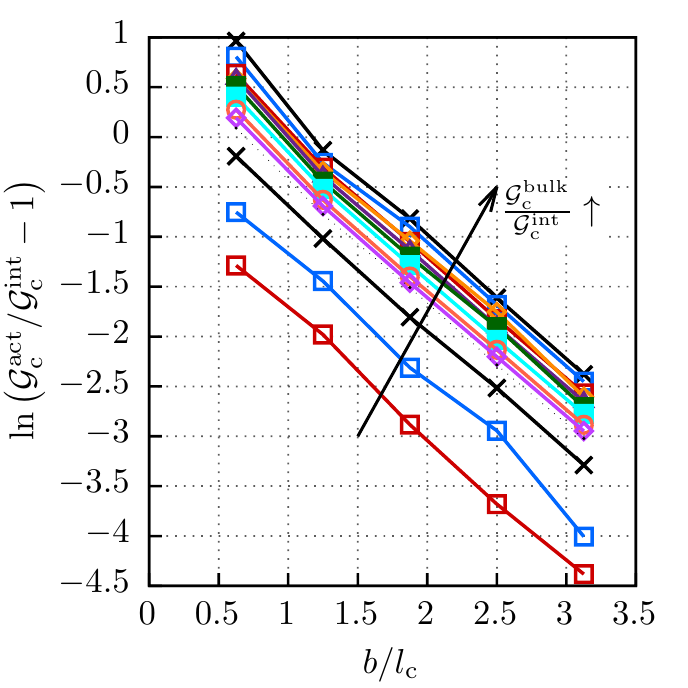}
	\label{fig:largeratb}}\\
	\vspace{0.3cm}
	\footnotesize 
	\begin{tabular}{rl|l|l|l|l}
	Legend for (a):\hspace{-0.3cm} &\legpoint{2} $b/l_\text{c}\equiv 0.625$ & \legpoint{3} $b/l_\text{c}\equiv 1.25$ & \legpoint{4} $b/l_\text{c}\equiv 1.875$ &  \legpoint{5} $b/l_\text{c}\equiv 2.5$ &\legpoint{6} $b/l_\text{c}\equiv 3.125$\\    
	\end{tabular}
	\caption{Influence of large ratios $\mathcal{G}_\text{c}^\text{bulk}/\mathcal{G}_\text{c}^\text{int}$: (a) For five ratios $b/l_\text{c}$, larger ratios $\mathcal{G}_\text{c}^\text{bulk}/\mathcal{G}_\text{c}^\text{int}$ have been investigated. Both, the ratios from the regularisation length scales and from the fracture toughnesses, have a nonlinear influence on the exaggeration $g=\mathcal{G}_\text{c}^\text{act}/\mathcal{G}_\text{c}^\text{int}$. The black dashed line is the limit exaggeration which is reached for $b/l_\text{c}=0$, \textit{i.e.} the interface completely vanishes and $\mathcal{G}_\text{c}^\text{act}=\mathcal{G}_\text{c}^\text{bulk}$. (b) The results presented in (a) are plotted over $b/l_\text{c}$ analogue to Fig.~\ref{fig:lca_sim}. Despite the large ratios, the curves are still only vertically shifted, which verifies the assumption for the exponential fits in Fig.~\ref{fig:lca_sim}. It can be seen, that the nonlinearity leads to a vanishing vertical shift of different curves for larger ratios. From the bottom to the top, the ratio $\mathcal{G}_\text{c}^\text{bulk}/\mathcal{G}_\text{c}^\text{int}$ increases from 1.5 to 20 and takes values according to the abscissa in (a).  }
	\label{fig:schemWorklargerat}
\end{figure}

\subsection{Compensation procedure}
To obtain crack propagation at the physical value $\mathcal{G}_\text{c}^\text{int}$, the interface fracture toughness is lowered to a new input value $\hat{\mathcal{G}}_\text{c}^\text{int}$. This new value has to satisfy
\begin{equation}
\label{eq:lca_exaggeration}
\hat{\mathcal{G}}_\text{c}^\text{int}\cdot g\left(  \frac{\mathcal{G}_\text{c}^\text{bulk}}{\hat{\mathcal{G}}_\text{c}^\text{int}},\frac{b}{l_\text{c}} \right)=\mathcal{G}_\text{c}^\text{act}=\mathcal{G}_\text{c}^\text{int}
\end{equation}
in order to properly compensate for the influence of the fracture toughness of the bulk material. Rerarranging 
Equation~\eqref{eq:lca_exaggeration} yields
\begin{equation}
\label{eq:lca_compensationfactor}
\frac{\hat{\mathcal{G}}_\text{c}^\text{int}}{\mathcal{G}_\text{c}^\text{int}}=\left[ g\left(  \frac{\mathcal{G}_\text{c}^\text{bulk}}{\hat{\mathcal{G}}_\text{c}^\text{int}},\frac{b}{l_\text{c}} \right) \right]^{-1}\coma
\end{equation}
where the right hand-side can be identified as the compensation factor. It is noted that Equation~\eqref{eq:lca_compensationfactor} 
cannot be solved directly because the exaggeration itself depends on~$\hat{\mathcal{G}}_\text{c}^\text{int}$. This can be overcome 
by an interpolation between the results given in Figure~\ref{fig:lca_sim}.

Figure~\ref{fig:compensation_large} presents sets of curves for 
$\mathcal{G}_\text{c}^\text{bulk}/\mathcal{G}_\text{c}^\text{int}\equiv \text{const}$. For a given ratio $b/l_\text{c}$ the factor 
for lowering the interface fracture toughness is given by the ordinate. The curves are as expected: for large ratios of
$b/l_\text{c}$ there is no need for compensation, whereas larger ratios 
$\mathcal{G}_\text{c}^\text{bulk}/\mathcal{G}_\text{c}^\text{int}$ call for smaller compensation factors when 
\mbox{$b/l_\text{c}\rightarrow 0$}. Another interesting aspect is the fact, that the curves in Figure~\ref{fig:compensation_small} 
seem to converge towards a limiting case. The existence of a limiting compensation factor is reasonable, because the compensated 
fracture toughness would otherwise hit negative values eventually. 

\subsection{Larger ratios of $\mathcal{G}_\text{c}^\text{bulk}/\mathcal{G}_\text{c}^\text{int}$ and limitations of the approach}
For five different ratios $b/l_\text{c}=\lbrace0.625,1.25,1.875,2.5,3.125\rbrace$, larger ratios of $\mathcal{G}_\text{c}^\text{bulk}/\mathcal{G}_\text{c}^\text{int}=\lbrace1.5\dots20 \rbrace$ have been investigated. The corresponding exaggeration for every set of parameters is given in Fig.~\ref{fig:schemWorklargerat}. Firstly, as already observed above, a variation of the fracture toughness ratio yields a nonlinear response of the exaggeration, which gets clearer when looking at large ratios $\mathcal{G}_\text{c}^\text{bulk}/\mathcal{G}_\text{c}^\text{int}$. The nonlinearity is reflected in Fig.~\ref{fig:largeratb}, where the curve families for larger ratios $\mathcal{G}_\text{c}^\text{bulk}/\mathcal{G}_\text{c}^\text{int}$ move more and more together. The exponential fits analogue to Fig.~\ref{fig:lca_sim} are excluded for the sake of readability. Secondly, large ratios $b/l_\text{c}$ reduce the exaggeration which can be seen from Fig.~\ref{fig:largerat}. This is in line with Fig.~\ref{fig:lca_sim}. \par 
The computation of such large fracture toughness ratios showed a limitation of the approach: Due to the dramatic decrease of the fracture toughness within the interface, convergence was hard to reach, which made a local mesh refinement and smaller load increments necessary. This could have been avoided -- at least partially -- when using a smooth transition from the bulk to the interface material, which raises the need for alternative interface regularisations. The following selection of numerical examples demonstrates the general functionality of the interface model. 
%%%%%%%%%%%%%%%%%%%%%%%%%%%%%%%%%%%%%%%%%%%%%%%%%%%%%%%%%%%%%%%%%%%%%%
\section{Numerical Examples}
\label{sec:numeric}
\subsection{Crack propagation along interface aligned with mesh}
\label{sec:numeric_conform}

\begin{figure}[tb]
	\centering
	\subfloat[Not compensated]{
		\label{fig:res_notcorr}
		\includegraphics[trim={0cm 0cm 0cm 0cm},clip]{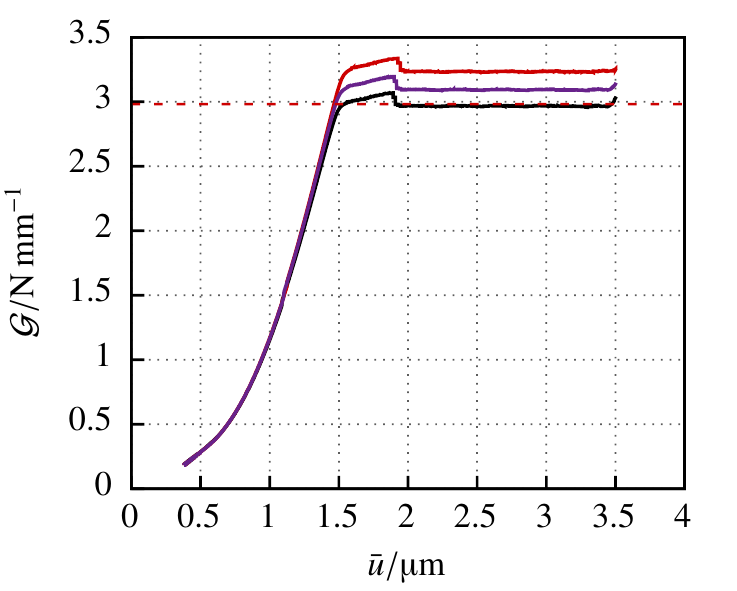}
	}
	\subfloat[Compensated]{
		\label{fig:res_corr}
		\includegraphics[trim={0cm 0cm 0cm 0cm},clip]{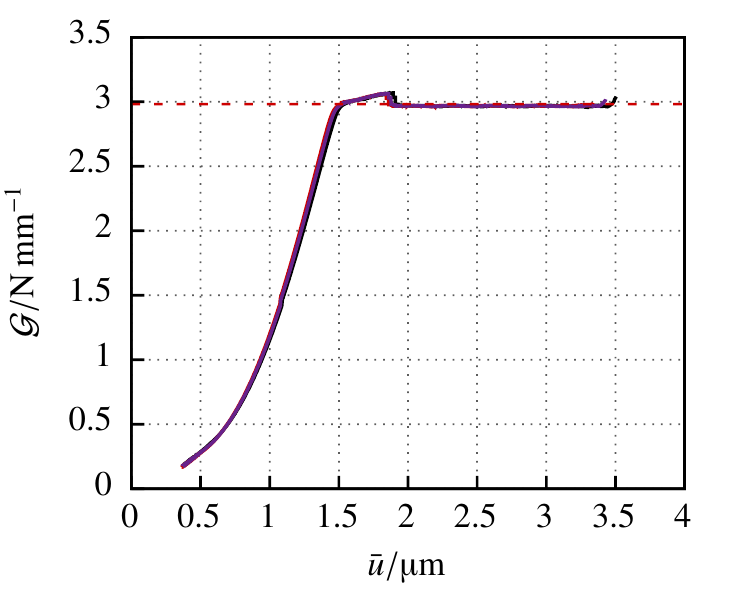}
	}
	\\
	\vspace{0.3cm}
	\footnotesize 
	\begin{tabular}{ l| l| l|| l}
		\legline{1} Reference &	\legline{3} $b/l_\text{c}=1.25$ & \legline{5} $b/l_\text{c}=2.08$ &  \legline{12} $\mathcal{G}_\text{c,num}^\text{int}$   \\
	\end{tabular}	
	
	\caption[Local correction approach -- energy release rate]{$\mathcal{G}$ for two ratios $b/l_\mathrm{c}$ with and without compensation are compared to the reference calculation with $b/l_\mathrm{c}\rightarrow\infty$. (a) The actual numerical energy release rate $\mathcal{G}_\text{num}^\text{act}$ for crack propagation increases for smaller ratios $b/l_\text{c}$. (b) The crack propagates at equal $\mathcal{G}$ for each compensated case. As expected, $\mathcal{G}$ recovers the value of the numerical interface fracture toughness $\mathcal{G}_\text{c,num}^\text{int}$, \textit{i.e.} $\mathcal{G}_\text{num}^\text{act}=\mathcal{G}_\text{c,num}^\text{act}=\mathcal{G}_\text{c,num}^\text{int}$. 
	}
	\label{fig:first_num}
\end{figure}

As a first example, the compensation procedure is applied to two cases used in the parameter study. For the values 
$\mathcal{G}_\text{c}^\text{int}=\SI{2.7}{\newton\per\milli\meter}$ and 
$\mathcal{G}_\text{c}^\text{bulk}=\SI{5.4}{\newton\per\milli\meter}$, and the ratios $b/l_\text{c}=\lbrace1.25,2.08\rbrace$ with 
$l_\text{c}=\SI{15}{\micro\meter}$, the results of the compensation are shown in Figure~\ref{fig:compensation_small} by means of red and purple arrows. Since 
$\mathcal{G}_\text{c}^\text{bulk}/\mathcal{G}_\text{c}^\text{int}=2$, the intersections of two vertical lines (red and purple 
arrows) at $b/l_\text{c}=\lbrace1.25,2.08\rbrace$ with the orange curve are needed. From these intersections, two horizontal lines 
give the ordinate and thus, the compensation factors 
$\hat{\mathcal{G}}_\text{c}^\text{int}/\mathcal{G}_\text{c}^\text{int}\approx\lbrace0.911,0.958\rbrace$ for the given cases. 
Instead of the original value for the interface fracture toughness $\mathcal{G}_\text{c}^\text{int}$, the corrected interface 
fracture toughness $\hat{\mathcal{G}}_\text{c}^\text{int}$ is applied together with the bulk fracture toughness 
$\mathcal{G}_\text{c}^\text{bulk}$.

Figure~\ref{fig:first_num} presents the energy release rates before and after correction. Both cases are compared with the 
reference case. The compensation has the correct effect and eliminates the bulk influence. 

\begin{figure}[!t]
	\centering
	\subfloat[Geometry and boundary conditions ]{
		\label{fig:geom_BC_case2}
		\includegraphics[trim={0cm 0cm 0cm 0cm},clip]{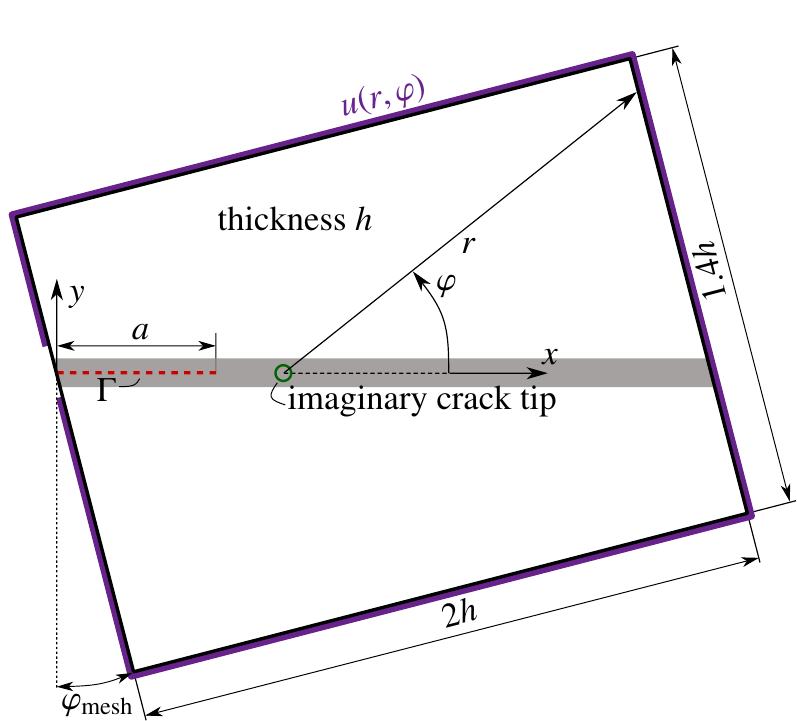}
	}%\hspace{1cm}
	\subfloat[Initial notch and prerefined mesh for $\varphi_\text{mesh}=\SI{15}{\degree}$]{
		\label{fig:geom_iniPF_case2}
		\raisebox{0.45cm}{\includegraphics[trim={0cm 0cm 0cm 0cm},clip]{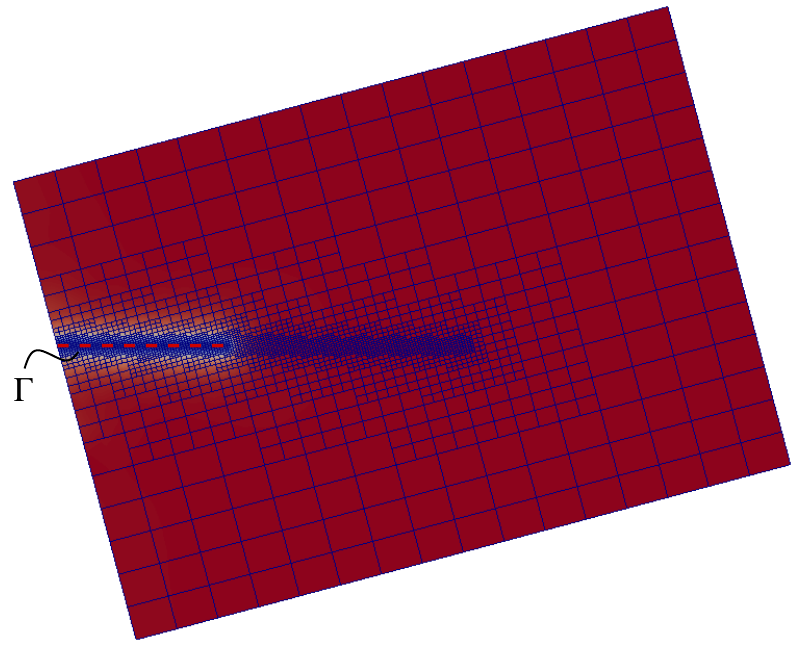}}
	}
	\caption[]{Geometry and boundary conditions: (a) The geometry with $h=2a=\SI{1}{\milli\meter}$ is described. As for the parameter study, surfing boundary conditions  $u(r,\varphi)$ of a mode-I crack are assigned to all nodes on the purple edges. The grey stripe resembles the diffuse interface. (b) The predefined notch is initialised by setting the phase-field $c=0$ (blue) along the red dashed line. The interface half width is set to $b=\SI{25}{\micro\meter}$, the length scale to $l_\text{c}=\SI{15}{\micro\meter}$.}
	\label{fig:geom_BC_iniPF_case2}
	
	\includegraphics[trim={0cm 0cm 0cm 0cm},clip]{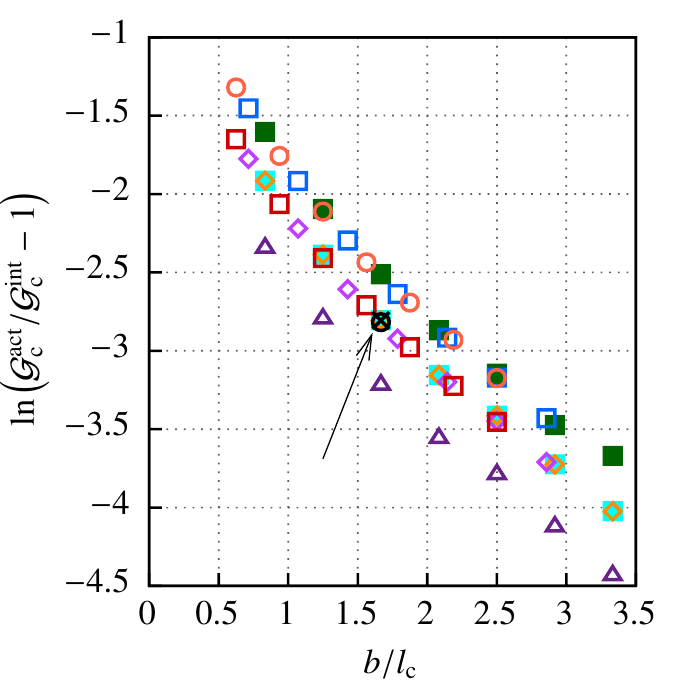}
	\caption[]{Exaggeration for arbitrary mesh orientation: The black arrow points at the exaggeration data (black circle/cross) which has been extracted from the simulations with $\varphi_\text{mesh}=\lbrace15,20\rbrace\,\si{\degree}$. The mesh seems to be sufficiently fine in the vicinity of the abrupt change in the fracture toughness since the points perfectly match the data from the parameter study. For details on colouring, the reader is referred to Figure~\ref{fig:lca_sim}.}
	\label{fig:exag_case2}
\end{figure}
\subsection{Crack propagation along interface not aligned with mesh}
\label{sec:numeric_nonconform}

In the simulations that have been presented up to this point, the interfaces were aligned with the given mesh and the interface 
half width $b$ was chosen such that it was a multiple of the element edge $\Delta h$. Of course, the orientation of the interface 
within the mesh can be arbitrary. Therefore, the second example features an interface which is inclined with the mesh lines. 
The geometry and the mesh are depicted in Figure~\ref{fig:geom_BC_iniPF_case2}. Again, surfing boundary conditions have been 
applied to all edges. Two orientation angles $\varphi_\text{mesh}=\lbrace15,20\rbrace\,\si{\degree}$ have been investigated. The 
\textsc{Young}'s modulus $E=\SI{210}{\giga\pascal}$ and the \textsc{Poisson}'s ratio $\nu=0.3$ are the same as in the previous 
simulations. Plane strain conditions have been adopted. 

For both cases, a reference calculation with a homogeneous fracture toughness 
$\mathcal{G}_\text{c}^\text{int}=\SI{2.7}{\newton\per\milli\meter}$ has been carried out. In Figure~\ref{fig:exag_case2}, the 
exaggeration functions from Figure~\ref{fig:lca_sim} are recapped and the results for different angles $\varphi_\text{mesh}$ are 
added (marked by a black arrow). Even though the numerical integration near the (abrupt) change in the fracture toughness may be 
slightly inaccurate due to the distorted mesh, the ratios perfectly match the results where the numerical integration errors have 
been eliminated. Thus, the model allows for arbitrary mesh orientations as long as the discretisation near the interface with a 
crack is sufficiently fine. 

\begin{figure}[!t]
	\centering
	\subfloat[Deflection/penetration relationship for linear elastic fracture mechanics~\cite{he_crack_1989}]{
		\includegraphics[trim={0cm 0cm 0cm 0cm},clip]{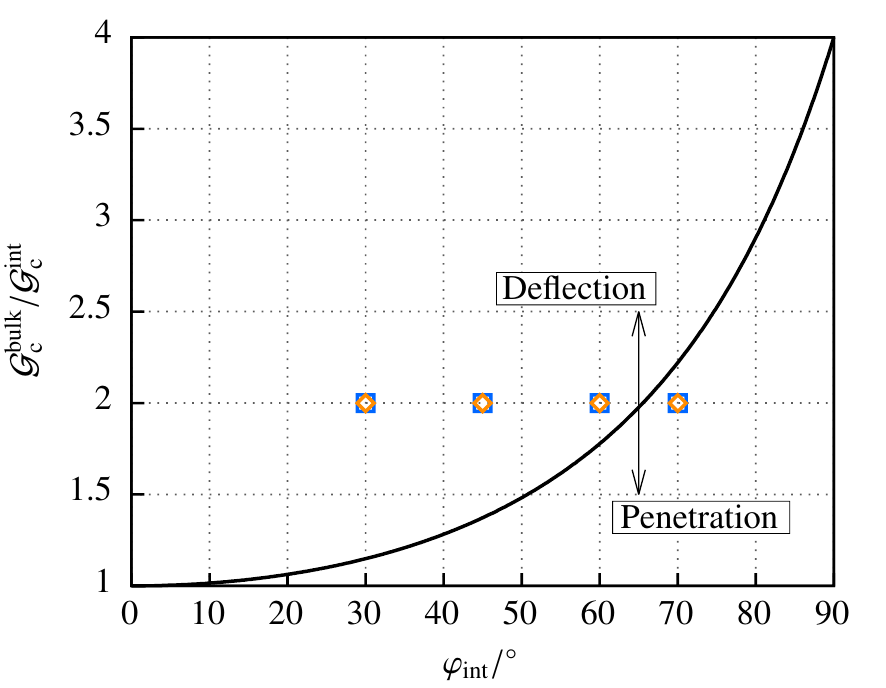}
		\label{fig:angle_deflection}
	}
	\subfloat[Crack length along inclined interface]{
		\includegraphics[trim={0cm 0cm 0cm 0cm},clip]{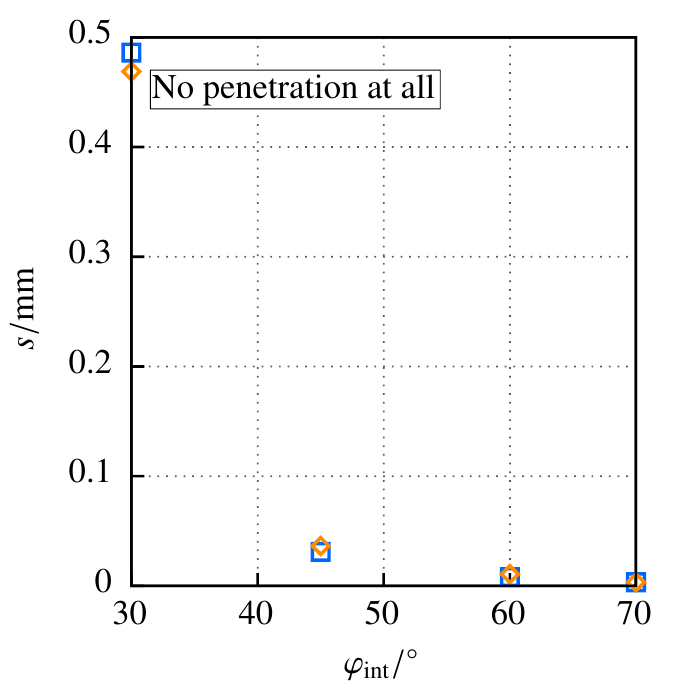}
		\label{fig:angle_following}
	}
	\\
	\vspace{0.3cm}
	\footnotesize 
	\begin{tabular}{ l| l}
		\legpoint{2} $\hat{\mathcal{G}}_\text{c}^\text{int}=\SI{2.459}{\newton\per\milli\meter}$ \& $b/l_\text{c}=1.25$ &	\legpoint{4} $\hat{\mathcal{G}}_\text{c}^\text{int}=\SI{2.586}{\newton\per\milli\meter}$ \& $b/l_\text{c}=2.08$ 
	\end{tabular}
	\caption{Relation between deflection and penetration for an inclined interface: (a)~Analytic relationship according to linear elastic fracture mechanics. The black curve is given in Equation~\eqref{eq:imping_rel} and describes the limiting case between penetration and deflection for the given parameters. It is noted that the interface fracture toughness $\mathcal{G}_\text{c}^\text{int}$ is achieved by assigning the compensated interface fracture toughness $\hat{\mathcal{G}}_\text{c}^\text{int}$ in the simulations. (b)~For different inclination angles the measure $s$ prescribes how long the crack follows the interface before penetrating into the opposite bulk material. Contour plots for the simulations represented by the orange diamond can be found in Figures~\ref{fig:mesh_case3_b} -- e.}
	\label{fig:angle_case3}	
\end{figure}

\subsection{Crack impinging on an interface}
\label{sec:numeric_impinging}
In linear elastic fracture mechanics, a crack impinging on an interface which is inclined compared to the crack path, has been 
considered analytically in~\cite{he_crack_1989}. Depending on the fracture toughness ratio 
$\mathcal{G}_\text{c}^\text{bulk}/\mathcal{G}_\text{c}^\text{int}$ and the interface inclination angle $\varphi_\text{int}$, 
a crack impinging on an interface is either deflected, or penetrates into the bulk 
material~\cite{paggi_revisiting_2017,he_crack_1989}:
\begin{equation}
\label{eq:imping_rel}
\frac{\mathcal{G}^\text{int}}{\mathcal{G}^\text{bulk}}=\frac{1}{16}\left\lbrace 
\left[3\cos\left(\frac{\varphi_\text{int}}{2}\right)+\cos\left(3\frac{\varphi_\text{int}}{2}\right) \right]^2+
\left[\sin\left(\frac{\varphi_\text{int}}{2}\right)+\sin\left(3\frac{\varphi_\text{int}}{2}\right)\right]^2
\right\rbrace\point
\end{equation}
For a ratio
\begin{equation}
\frac{\mathcal{G}_\text{c}^\text{int}}{\mathcal{G}_\text{c}^\text{bulk} }<\frac{\mathcal{G}^\text{int}}{\mathcal{G}^\text{bulk}}\coma
\end{equation}
the crack is deflected~\cite{he_crack_1989}. Otherwise, it penetrates into the bulk. The limiting curve between deflection 
and penetration is plotted in Figure~\ref{fig:angle_deflection}. It is noted that the inverse of Equation~\eqref{eq:imping_rel} is 
shown in line with the other ratios presented herein. The limiting curve has to be understood as a tendency of what will happen, 
rather than a prediction.

\begin{figure}[!t]
	\centering
	\subfloat[Initial phase-field]{
		\label{fig:mesh_case3_a}
		\includegraphics[trim={4.6cm 1.35cm 4cm 1.35cm},clip,scale=0.738]{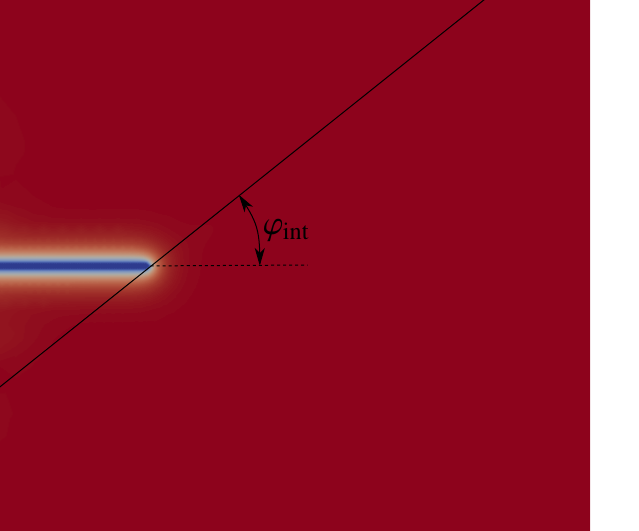}
	}
	\subfloat[$\varphi_\text{int}=\SI{30}{\degree}$]{
		\label{fig:mesh_case3_b}
		\includegraphics[trim={15.6cm 5cm 27cm 5cm},clip,scale=0.22]{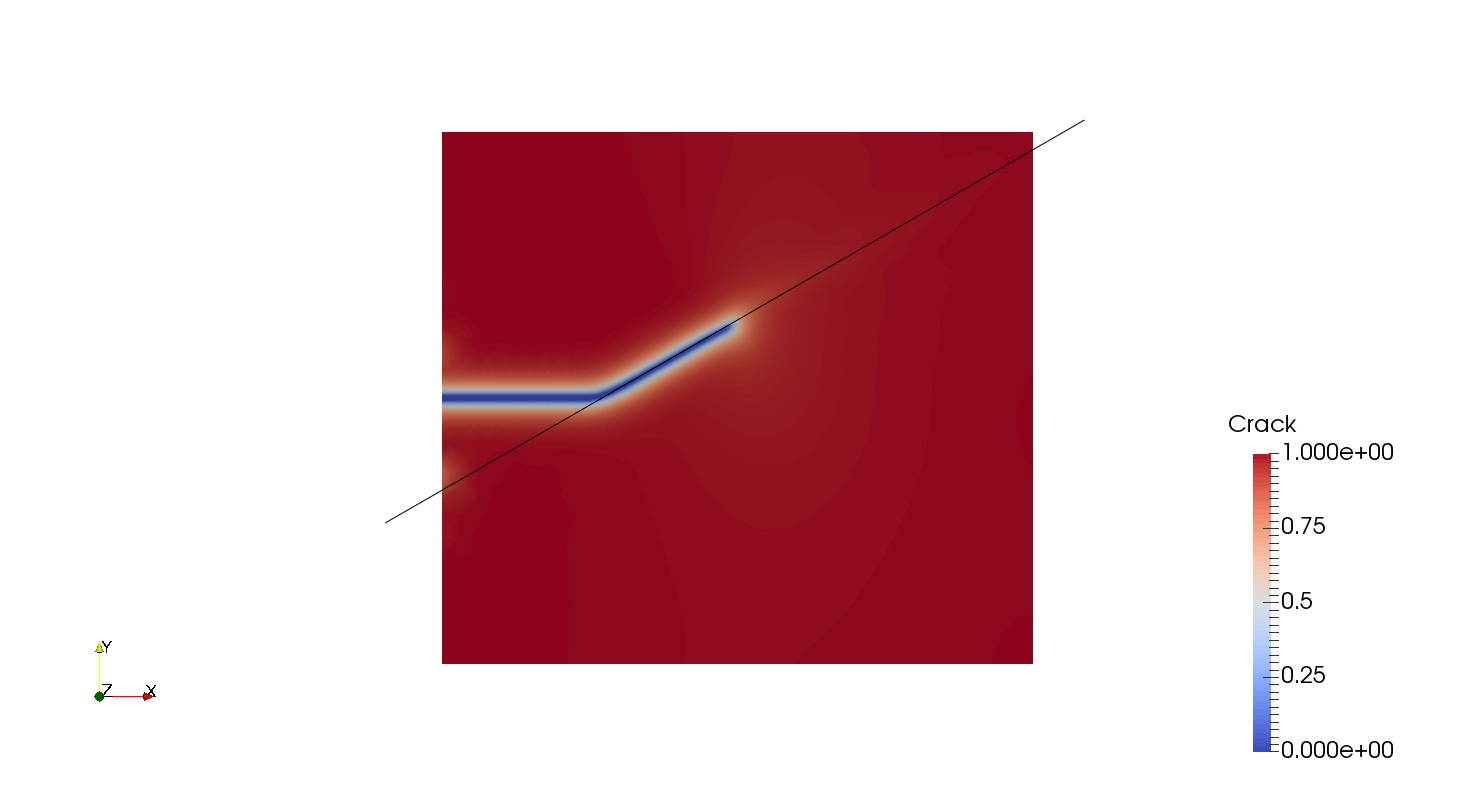}
	}
	\subfloat[$\varphi_\text{int}=\SI{45}{\degree}$]{
		\label{fig:mesh_case3_c}
		\includegraphics[trim={15.6cm 5cm 27cm 5cm},clip,scale=0.22]{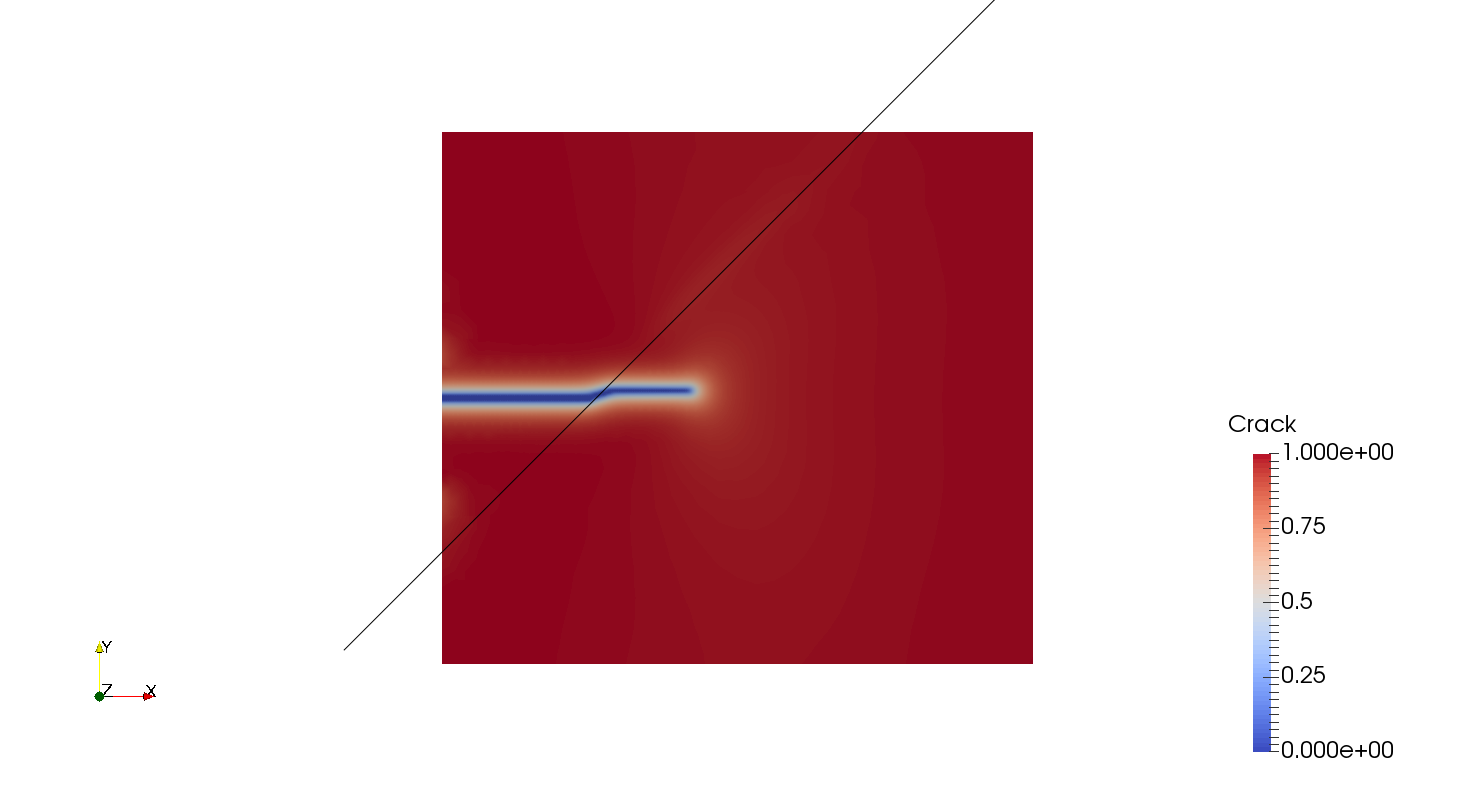}
	}
	\subfloat[$\varphi_\text{int}=\SI{60}{\degree}$]{
		\label{fig:mesh_case3_d}
		\includegraphics[trim={15.6cm 5cm 27cm 5cm},clip,scale=0.22]{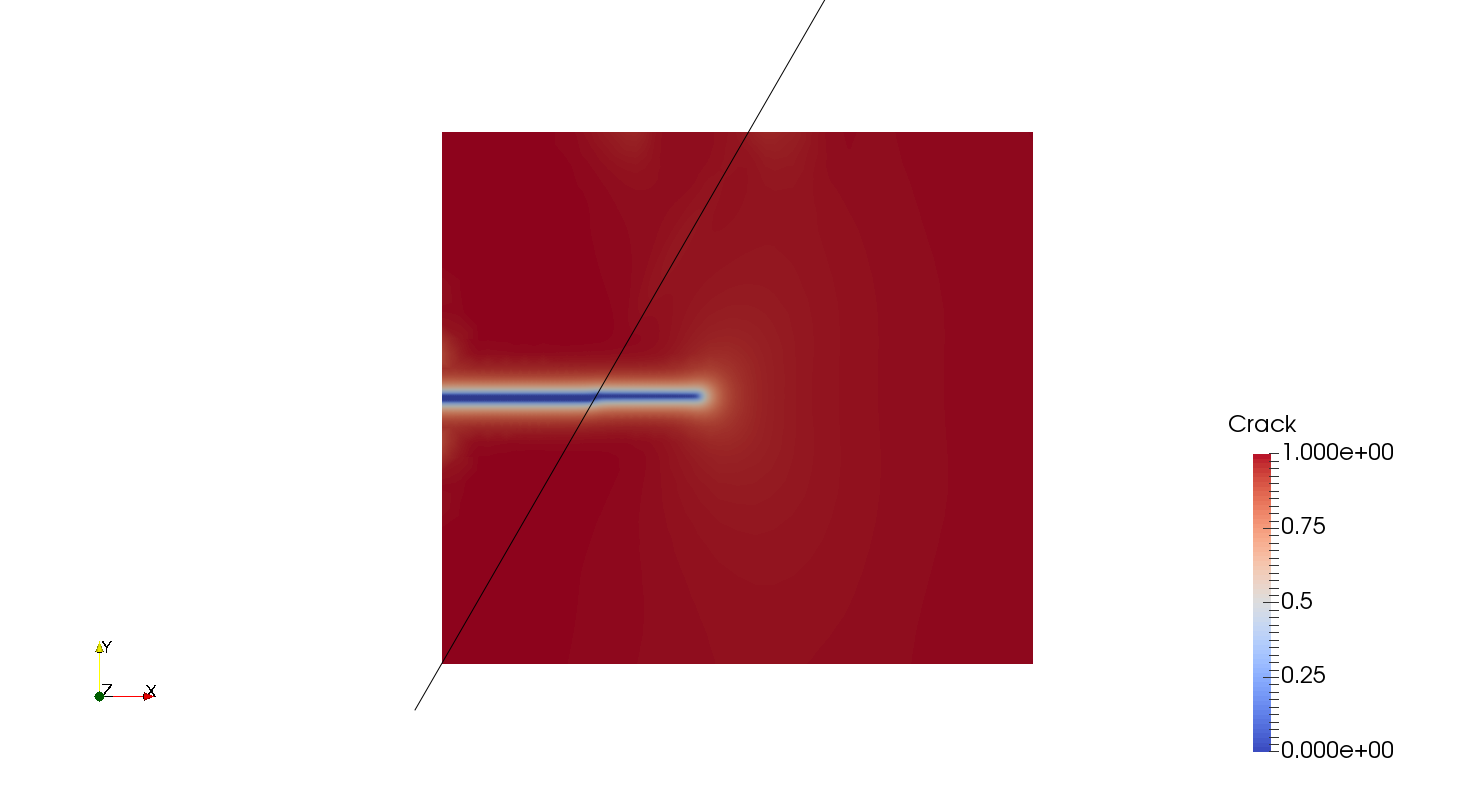}
	}
	\subfloat[$\varphi_\text{int}=\SI{70}{\degree}$]{
		\label{fig:mesh_case3_e}
		\includegraphics[trim={15.6cm 5cm 27cm 5cm},clip,scale=0.22]{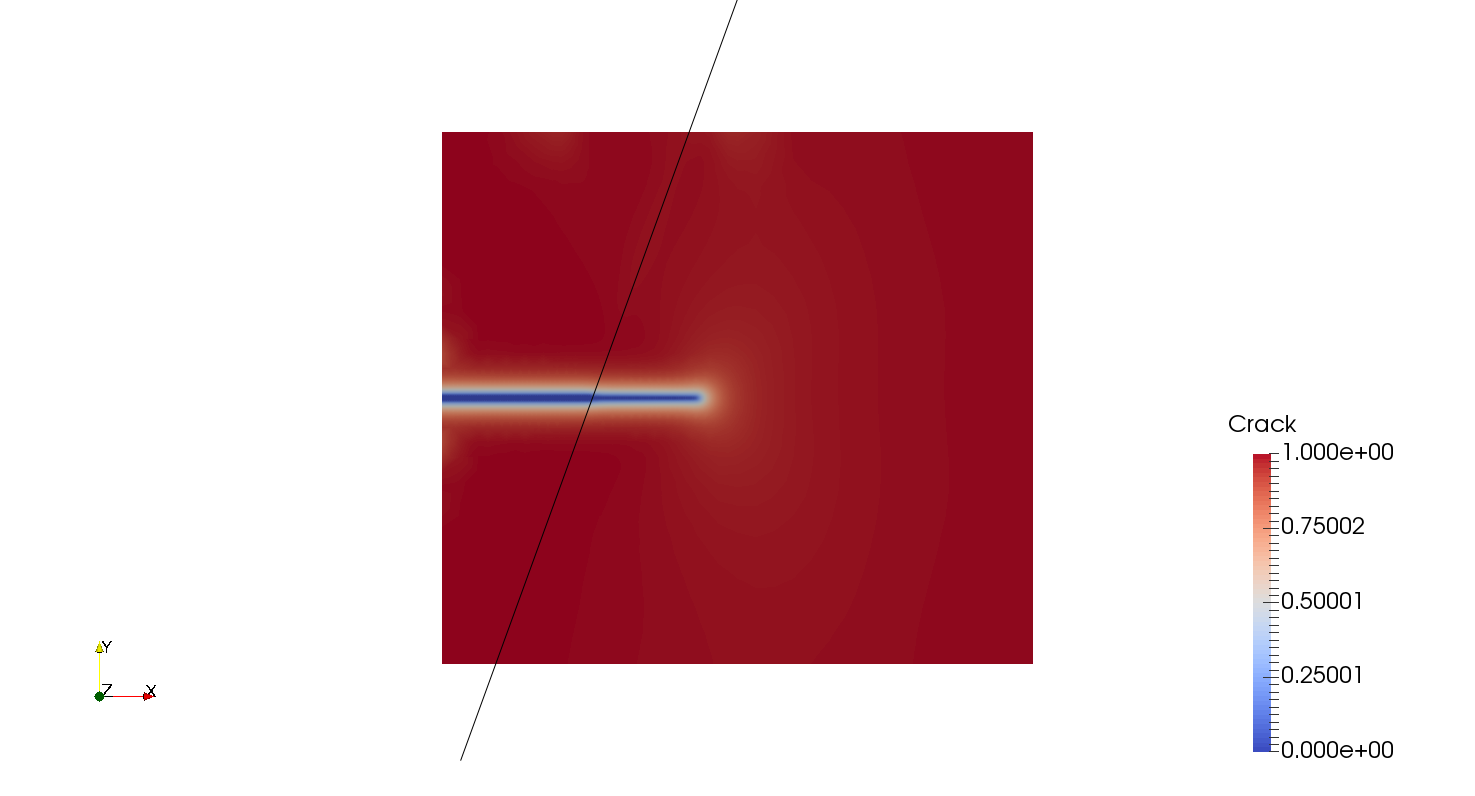}
	}
	\caption{Geometry and results for different inclination angles: (a) The initial phase-field and the inclination of the interface are depicted. The material parameters are chosen similar to the previous examples. A surfing boundary condition is applied similar to Figure~\ref{fig:geom_BC}. (b) -- (e) The resulting phase-fields for different inclination angles $\varphi_\text{int}$ (black lines) and $b/l_\text{c}=2.08$ are plotted. It can be seen that there is absolutely no deflection for $\varphi_\text{int}=\SI{70}{\degree}$. Starting from $\varphi_\text{int}=\SI{60}{\degree}$ deflection occurs. This is consistent with Figure~\ref{fig:angle_deflection} where the transition from deflection to penetration happens between these inclination angles.}
	\label{fig:mesh_case3}	
\end{figure}

For the two cases in Section~\ref{sec:numeric_conform} simulations for different inclination angles have been carried out to 
demonstrate the capability of the model. The different configurations are included in Figure~\ref{fig:angle_deflection}. The 
geometry is similar to Figure~\ref{fig:geom_BC} except for the domain size in the $y$-direction, which is enlarged by a factor of 
1.8 in a symmetric manner. The orientation of the inclined interface is shown in Figure~\ref{fig:mesh_case3_a}. Surfing boundary 
conditions are applied as well as plane strain conditions. It is noted that the compensated interface fracture toughness, 
$\hat{\mathcal{G}}_\text{c}^\text{int}$, is used to satisfy Equation~\eqref{eq:lca_exaggeration}, as also discussed in 
Section~\ref{sec:numeric_conform}. Adaptive local refinement has been used in all the simulations.

Figure~\ref{fig:angle_following} shows how long the crack follows the interface before it penetrates into the bulk material. 
For $\varphi_\text{int}=\SI{30}{\degree}$ no penetration occurred. For larger inclination angles, the crack started to penetrate 
the bulk after being deflected. As expected, no deflection occurred for $\varphi_\text{int}=\SI{70}{\degree}$, which is consistent 
with the analytical result. There is no significant difference in the results for the two values for $b/l_\text{c}$. The contour 
plots in Figure~\ref{fig:mesh_case3} can be compared to those obtained for a brittle interface in~\cite{paggi_revisiting_2017}, Figure~9. The crack lengths following the interface match well qualitatively. Differences occur for the cracking behaviour after 
penetrating into the bulk material. Different from here,  a cohesive-zone model is introduced in~\cite{paggi_revisiting_2017} 
which influences the direction of the crack after penetrating into the bulk material. The crack does not propagate horizontally 
as in Figures~\ref{fig:mesh_case3_b} -- e. From Figures~\ref{fig:mesh_case3_b} -- e, which show the resulting phase-field for 
$b/l_\text{c}=2.08$, it also becomes clear that adaptive refinement is most useful for this kind of phase-field simulations. 
Without adaptive refinement, the complete region below the interface and above the $x$-axis would have had to be refined a priori.

%%%%%%%%%%%%%%%%%%%%%%%%%%%%%%%%%%%%%%%%%%%%%%%%%%%%%%%%%%%%%%%%%%%%%%
\section{Conclusions and Outlook}
\label{sec:conclusion}
In this work, a phase-field model has been introduced for interface failure. The standard phase-field model can describe bulk 
failure, but is not able to describe adhesive interfacial failure without a modification. Herein, the interface was distributed
over a finite length and assigned an interface fracture toughness $\mathcal{G}_\text{c}^\text{int}$. 

An interaction between the width of the interface and the characteristic length scale of the phase-field model for the bulk can 
occur for certain ratios of the smearing widths of interface and the bulk fracture zones. For one-dimensional simulations, an analytical
expression was derived to compensate for this effect. For two dimensions, a numerical correction approach was developed motivated 
by a theoretical reasoning and several simulations which suggest that a compensation, herein named exaggeration function, can be 
described by an exponential decay. Numerical simulations underpinned this assumption. Moreover, it was shown that the exaggeration 
function is independent of absolute values of the fracture toughnesses of the interface or the bulk, as it only depends on ratios 
which makes the relation universal. 

Numerical examples demonstrate the basic functionality of the compensation approach and the applicability to arbitrary mesh 
orientations, which is a main feature and advantage over models which exploit interface elements. Finally, a crack impinging on an 
interface was simulated. The results compare well with analytical relations from linear elastic fracture mechanics. 

\section*{Acknowledgements}
The authors gratefully acknowledge support by the Deutsche Forschungsgemeinschaft in the Priority Program 1748 “Reliable simulation techniques in solid mechanics. Development of non-standard discretisation methods, mechanical and mathematical analysis” under the project KA3309/3-2, and from the European Research Council under Advanced Grant PoroFrac (grant number 664734).

%%%%%%%%%%%%%%%%%%%%%%%%%%%%%%%%%%%%%%%%%%%%%%%%%%%%%%%%%%%%%%%%%%%%%%



\bibliographystyle{elsarticle-num} 
\bibliography{literatur.bib}


%% \begin{thebibliography}{00}


%% \bibitem{testbibitem}

%% \end{thebibliography}
\end{document}